\documentclass[traditabstract]{aa} 

\usepackage{natbib}
\bibpunct{(}{)}{;}{a}{}{,} 
\usepackage{color}
\usepackage[pdftex]{graphicx}
\usepackage{txfonts}

\usepackage{epstopdf}

\begin{document}
   \title{Mid-infrared properties of nearby low-luminosity AGN at high angular resolution\thanks{Based the ESO observing programs 083.B-0536 and 086.B-0349}}


   \author{D. Asmus
          \inst{1 ,2}\fnmsep\thanks{\email{asmus@astrophysik.uni-kiel.de}}
          \and
           P. Gandhi\inst{3}
          \and
          A. Smette\inst{1}
          \and
          S. F. H\"onig\inst{4}
          \and
          W. J. Duschl\inst{2,5}
           }

   \institute{European  Southern Observatory, Casilla 19001, Santiago 19, Chile
         \and
             Institut f\"ur Theoretische Physik und Astrophysik,
             Christian- Albrechts-Universit\"at zu Kiel, Leibnizstr. 15, 24098 Kiel Germany
         \and
             Institute of Space and Astronautical Science (ISAS), Japan Aerospace Exploration Agency, 3-1-1 Yoshinodai, chuo-ku, Sagamihara, Kanagawa 252-5210, Japan
         \and
             Department of Physics, University of California in Santa Barbara, Broida Hall, Santa Barbara, CA 93106-9530, USA
         \and
             Steward Observatory, The University of Arizona, 933 N. Cherry Ave, Tucson, AZ 85721, USA
              }

   \date{Received April XX, 2010; accepted June XX, 2010}

 
  \abstract
   {
In this work high spatial resolution mid-infrared (MIR) 12\,$\mu$m continuum imaging of low-luminosity active galactic nuclei (LLAGN) obtained by VLT/VISIR is presented.
The goal of this investigation is to determine if the nuclear MIR emission of LLAGN is consistent with the existence of a dusty obscuring torus, the key component of the unification model for AGN. 
Based on available hard X-ray luminosities and the previously-known tight correlation between the hard X-ray and 12\,$\mu$m luminosities, a sample of 17 nearby LLAGN without available VISIR N-band photometry was selected. 
Combined with archival VISIR data of 9 additional LLAGN with available X-ray measurements, the dataset represents the bulk of southern LLAGN currently detectable from the ground in the MIR.
Of the 17 observed LLAGN, 7 are detected, while upper limits are derived for the 10 non-detections. 
Thus, the total number of AGN detected with VLT/VISIR increases to more than 50. 
All detections except NGC 3125 appear point-like  on a spatial scale of $\sim 0.35\arcsec$.
The detections do not significantly deviate from the known MIR-X-ray correlation but extend it by a factor of $\sim 10$ down to luminosities $ < 10^{41}$\,erg/s with a narrow scatter ($\sigma = 0.35\,$dex, Spearman Rank $\rho = 0.92$ ).
The latter is dominated by the uncertainties in the X-ray luminosity. 
Interestingly, a similar correlation with comparable slope but with a normalization differing by $\sim 2.6$ orders of magnitude has been found for local starburst galaxies.
In addition, the VISIR data are compared with lower spatial resolution data from \textit{Spitzer}/IRS and \textit{IRAS}. 
By using a scaled starburst template SED and the PAH 11.3\,$\mu$m emission line the maximum nuclear star formation contamination to the VISIR photometry is restricted to $\lesssim 30\%$ for 75\% of the LLAGN. 
Exceptions are NGC 1097 and NGC 1566, which may possess unresolved strong PAH emission. 
Furthermore,  within the uncertainties the MIR-X-ray luminosity ratio is unchanged over more than 4 orders of magnitude in accretion rate. 
These results are consistent with the existence of the dusty torus in all observed LLAGN, although the jet or accretion disk as origin of the MIR emission cannot be excluded.
Finally, the fact that the MIR-X-ray correlation holds for all LLAGN and Seyferts makes it a very useful empirical tool for converting between the MIR and X-ray powers of these nuclei.
}



   \maketitle
%

\section{Introduction}
 Low-luminosity active galactic nuclei (LLAGN; $L_\mathrm{2-10keV}  \lesssim   
10^{42}$\,erg/s) have drawn increasing attention in recent years for various
reasons: it has been realized that this population is not only the most common
of all active galaxies (AGN) -- about two-thirds  of all AGN  are  LLAGN
\cite[see][]{ho_nuclear_2008} -- but they may also represent AGN in the initial
or final stages of their active phases. The underlying idea is that every galaxy
harbors a black hole (BH) in its center that grows and evolves  to a
supermassive BH during many phases of nuclear activity. The key components
describing AGN in the so called unification scheme
\citep{antonucci_unified_1993} such as the accretion disk and the dusty
obscuring torus very likely vanish during non-active phases and are built-up
again when new material reaches the galactic nucleus due to any kind of larger
scale disturbance. But it is unknown how the processes of formation and
evolution of both are connected. This is partially because the
physical nature and morphology of the disk and torus is still not exactly known.
However, high spatial resolution mid-infrared (MIR) observations made progress
in the modeling of the torus possible. They favor a clumpy distribution of
optically thick clouds with a small filling factor
\citep[e.g.][]{nenkova_dust_2002, dullemond_clumpy_2005, hoenig_radiative_2006,
schartmann_three-dimensional_2008, nenkova_agn_2008-1, hoenig_dusty_2010}.
Apart from the fact that the mechanism, which maintains the geometrical thickness of the torus,
is unknown, there are various theoretical works predicting its disappearance
below certain bolometric luminosities or accretion rates
\citep{elitzur_agn-obscuring_2006, hoenig_active_2007}.
But observational proof of this disappearance is still lacking.

The putative dusty torus of AGN is best observed in the MIR, where the thermal
emission from the warm dust is strongest. 
Ground-based MIR instruments like
VLT/VISIR \citep{lagage_successful_2004} are best suited to perform these observations due to their high spatial resolution.
And an increasing number of AGN studies demonstrate their power to isolate the AGN from surrounding
starbursts \citep[e.g.][]{gorjian_10_2004, galliano_mid-infrared_2005,haas_visir_2007, siebenmorgen_nuclear_2008,ramos_almeida_infrared_2009}. 
Furthermore, mid-infrared interferometric observations were able to resolve the dusty torus in nearby AGN and showed that it is only a few parsec in size \citep[e.g.][]{jaffe_central_2004,raban_resolving_2009,tristram_parsec-scale_2009, burtscher_dust_2009-1} and scales by the square root of the AGN luminosity \citep{tristram_size-luminosity_2011}.

On the other hand, past low-resolution MIR observations led to the discovery of a correlation
between the MIR  and X-ray luminosities \citep{krabbe_n-band_2001,
lutz_relation_2004}.
Here, the observed nuclear MIR emission is believed to represent the reprocessed UV/X-ray radiation from the dusty clouds inside the torus.
This conclusion is strongly supported by fitting of high-spatial resolution spectrophotometry with clumpy torus models \citep{hoenig_dusty_2010-1, alonso-herrero_torus_2011}. 

Then, recently gathered VISIR photometry enabled the best estimate of AGN emission around 12\,micron, free of stellar contamination, resulting in a strong correlation with the intrinsic X-ray 2-10\,keV luminosity \citep{horst_small_2006, horst_mid_2008, horst_mid-infrared_2009,
gandhi_resolving_2009},  which is well-described by a power-law ($L_\mathrm{MIR} \propto L_\mathrm{X}^b$) with a slope $b \sim 1$. 
 Similar results have also been found with other ground-based high spatial resolution MIR instruments
\citep[e.g.][]{levenson_isotropic_2009}. 
The luminosity correlation for the different AGN types is indistinguishable within the measured precision, and it can be used to e.g. assess the bolometric IR luminosity of AGN when no high resolution IR data is available \citep[e.g.][]{ mullaney_defining_2011}.  

However, the lowest luminosity objects of the observed samples \citep[in e.g.][ hereafter G+09]{gandhi_resolving_2009}, in particular the 3 LINERs, show a tendency to deviate from this correlation: all objects with an X-ray luminosity $L_{2-10\mathrm{keV}} < 10^{42}\,\mathrm{erg/s}$ exhibit a MIR-excess (or an X-ray deficit) of $\sim 0.3$\,dex.
The total number of included LLAGN  was too small, to give statistical significant evidence for a real deviation.
While such a deviation could indicate a change in the physical structure of the AGN at low luminosities, one first has to exclude other causes as e.g. significant MIR emission from dust heated by circum-nuclear star formation regions. 
In particular at low luminosities, the latter can be of comparable MIR brightness with respect to the AGN or even brighter. 
Furthermore, the proposed collapse of the dusty torus (see above) would lead to a deficit of MIR emission, opposite to the indicated deviation.
Instead, a separate AGN component would have to explain the MIR-excess (or X-ray-deficit) at low luminosities, if real.   
One candidate would be the jet, which is known to emit copious amounts of synchrotron emission over the whole observable wavelength range. 
In particular in LINERs (short for low-ionization nuclear emission-line region, a sub-class of LLAGN), the jet is assumed to be much more powerful than in Seyferts, explaining relative radio-brightness of these optically faint objects. 
For example, indications of significant jet contribution to the MIR emission of LLAGN was found in \cite{ hardcastle_active_2009}.  
In fact for NGC 4486, a well-known LINER, the observed MIR emission can be explained only with synchrotron emission of different jet components \citep{perlman_mid-infrared_2007}.  
 
However, the inner accretion disk might also emit MIR synchrotron emission, when in the so-called 
radiatively inefficient accretion flow (RIAF) or advection dominated
accretion flow (ADAF) mode.
The latter was theoretically predicted by \cite{narayan_advection-dominated_1994} to occur at low 
accretion rates:  the standard geometrically thin, optically thick disk \citep{shakura_black_1973} transforms
into a geometrically thick, optically thin accretion disk in the inner part. 
In addition, the thin accretion disk may still exist in the outer parts as a so-called truncated disk emitting thermal MIR radiation. 
Later, observations gave indeed evidence for this accretion mode in LLAGN, and thus it is favored for these sources \citep[see][ for an overview] {ho_nuclear_2008, yuan_advection-dominated_2007}. 
Of course a change in the accretion mode could also invoke a change in the UV or X-ray emission properties, which in course could by itself lead to an observed X-ray-deficit (or MIR-excess) compared to Seyferts accreting in a different mode.

To summarize, circum-nuclear star formation, the putative torus, a jet or the accretion disk, can all significantly contribute to the observed MIR emission of LLAGN.
By characterizing the MIR spectral energy distribution,  it might be possible to distinguish between these different possible components, or at least constrain their relative strength -- and thus test the underlying models for LLAGN. 
This work represents the first step by investigating whether the apparent deviation from the luminosity correlation of G+09 still holds for a larger sample of LLAGN, and finding  possible explanations for any
deviation. 
Therefore, new VISIR imaging data of a sample of 17 LLAGN is presented here and will be compared to the previous higher luminosity AGN sample from G+09.

The structure of this paper is as follows: First, the sample selection and
properties are described in Section~\ref{sec:sample}. The following section
(\ref{sec:obs}) details the VISIR observations, MIR data reduction and
photometric measurements. The obtained results are presented in
Sect.~\ref{sec:res}. And combined with additional objects from the literature
the MIR-X-ray correlation for LLAGN is investigated. Comparisons to
\textit{Spitzer}/IRS and \textit{IRAS} are presented and discussed in
Sect.~\ref{sec:dis}, as well as possible star formation contamination. In
addition, a possible dependence of the MIR-X-ray luminosity ratio on the
accretion rate is discussed in Sect.~\ref{sec:rat}.
The paper ends with a summary and conclusions (Sect.~\ref{sec:con}).

\section{Properties of the LLAGN sample}\label{sec:sample}
The sample of LLAGN was chosen with the goal of testing the previously
found MIR-excess/X-ray-deficit. Thus, the sample is not suited to detect objects
significantly above the correlation (i.e. showing a MIR deficit). In detail, 17
LLAGN were selected according to the following main
criteria: {\it i)} visible from Paranal, {\it ii)} a nuclear X-ray detection in
the 2-10\,keV band with a corrected $L_{2-10\mathrm{keV}} \lesssim
10^{42}$\,erg/s, {\it iii)} classified as LINER or Seyfert, and {\it iv)} an
expected flux $f_{12\,\mu\mathrm{m}} \ge 10\,\mathrm{mJy}$ assuming that the
objects follow the MIR-X-ray correlation as found by \cite{horst_mid_2008}
(9 objects) or that they follow the trend of MIR-excess (or
X-ray-deficit) seen for the 3 LINERS in the same work (9 objects). 
A flux of 10\,mJy is roughly the lowest unresolved 12\,$\mu$m flux VLT/VISIR can detect in
a narrow-band filter in a reasonable amount of execution time ($\sim$ 1\,h)
under average conditions. In addition, there are objects for which the selection
criteria apply but which have sufficient VISIR N-band imaging data available in
the literature (e.g. Centaurus A, M 87). These objects will be included in the
latter part of this analysis. Thus, the total sample comprises 26 LLAGN with 11
LINERs, 2 type 1 Seyferts and 13 type 2 Seyfert. 
In addition, NGC 1404 which is classified as a narrow emission-line galaxy
(NELG) was included in this sample because it was expected to be bright enough in the MIR.
In this work only the following main classes will be distinguished: type 1 (1.0-1.5), type 2 (1.8-2.0 and 1h, 1li) and LINERs. The AGN classification of \cite{veron-cetty_catalogue_2010} have been used in all cases where available (see table~\ref{tab:lum}). 
NGC 3125, NGC 5363, NGC 5813 and NGC 7626 are not in this catalog but have been classified as LINERs by \cite{carrillo_multifrequency_1999} and \cite{ho_search_1995}. 
Note that some objects have multiple different classifications (as either Seyfert or LINER) in the literature (e.g. NGC 1097 or NGC 7213). 
All these classifications are defined by optical properties (emission line ratios), which can be heavily affected by extinction. 
Furthermore, a significant number of AGN might even be completely obscured in the optical and thus remain unidentified as active \citep[e.g.][]{goulding_towards_2009}.

Additional notes on individual LLAGN included in this work, in particular concerning their X-ray properties, can be found in Appendix~\ref{app:prop}.

\subsection{Distances}
As the selected LLAGN are very nearby in a cosmological sense ($\sim 20$\,Mpc)
the motion due to local gravity potentials can dominate over the Hubble flow.
Thus, for most nearby galaxies it can be necessary to use redshift-independent
distance measurement methods. 
Here, three different distance collections are used: a) for comparison the redshift-based luminosity distances $D_L$ from NED
using redshifts corrected for the Earth motion relative to the CMB reference
frame with $H_0 = 73, \Omega_\mathrm{m} = 0.27,$ and $\Omega_\mathrm{vac} = 0.73
$, b) the distances given by \cite{tully_nearby_1988} employing the Tully-Fisher
(TF) relation, and c) distances of various methods with preference given to the most
precise and recent measurements, with the highest priority given to direct
Cepheid-based distance measurements and next using the surface brightness
fluctuation method (see \cite{jacoby_critical_1992} for a review and
comparison). If not available, the values based on the TF method or $D_L$ were
used. Note that there is no Tully-Fisher distance available for NGC 1667, NGC
3125 and
NGC 7626. References for the individual sources are listed in
Table.~\ref{tab:lum}. However, unless noted otherwise distance set c) will be
used.

\subsection{X-ray luminosities}\label{sec:xray}
The X-ray luminosities were selected from the literature and are
corrected for the distances adopted in each case. Preference was given to
the most recent observations with X-ray telescopes having a high angular
resolution, a large effective area, broad band energy coverage or the best
combination of these. The values used are
listed in Table~\ref{tab:lum} and represent in all cases the intrinsic
luminosities taking into account absorption. Therefore, they can of course
depend strongly on the spectral models used.
This affects mostly Compton-thick (CT, $N_{\rm H} > 1.5 \cdot 10^{24}$\,cm$^{-2}$) objects, and a number of LLAGN turned out to be such CT candidates. 
Their X-ray spectra are often dominated by scattered or extended emission and thus appear to be unabsorbed with a much lower luminosity than predicted by CT obscuration. 
Depending on the model assumed this leads to intrinsic 2-10\,keV luminosity estimates differing by orders of magnitude (e.g. for NGC 7743). 

To identify the ``true'' nature of the Compton-thick candidates, it is useful to compare the estimated X-ray luminosity to other intrinsic indicators, such as luminosities of the forbidden emission lines [OIII] $\lambda 5007\,$\AA\, and [OIV] $\lambda 25.89\,\mu$m. In \cite{panessa_x-ray_2006} a correlation between the 2-10\,keV and [OIII] luminosities has been found, which can be used to predict the intrinsic X-ray luminosity for the Compton-thick candidates. There is also evidence for significant absorption of [OIII] in the narrow line region \citep[NLR;][]{haas_spitzer_2005}. Whereas, the [OIV] is almost unabsorbed, making it a powerful tool \citep{melendez_new_2008}. \cite{diamond-stanic_isotropic_2009} present a large sample of local Seyferts with measured [OIV] and observed X-ray fluxes, whose luminosities correlate well for the 27 unabsorbed sources (Spearman rank $\rho = 0.84$ with $\log p = -7.3$). A bisector fit \citep{isobe_linear_1990} to these objects yields: 
\begin{equation}
  \log \left( \frac{L_{2-10\,\mathrm{keV}}}{\mathrm{erg\,s}^{-1}} \right) = -0.44 \pm 3.46 + (1.06 \pm 0.09) \left( \frac{L_\mathrm{[OIV]}}{\mathrm{erg\,s}^{-1}} \right)
\end{equation}
 Both luminosity correlations, X-ray-[OIII] and X-ray-[OIV], are used to predict the intrinsic X-ray luminosities of the CT candidates, and thus to distinguish between the Compton-thin or -thick scenario. 

In addition, it is well known that AGN can show huge X-ray variation, either
intrinsic or due to a change of obscuration/absorption in the innermost region
\citep[e.g.][]{murphy_monitoring_2007-1, risaliti_occultation_2007}.
To account for this an error of factor 2 (0.3\,dex) is assumed in all cases where
variation is evident or only one high spatial resolution X-ray observation is
available, while 0.6\,dex is applied for CT candidates. 
In particular, it is important to note that X-ray luminosities of the following
objects have been significantly revised since the sample was initially selected:
for NGC 5813 only \textit{ROSAT} data were available inferring a much higher
X-ray luminosity ($\log L_{2-10\mathrm{keV}} = 42.1$, \cite{schwope_rosat_2000})
than later found by \textit{Chandra} ($\log L_{2-10\mathrm{keV}} = 38.77$),
similarly for NGC 7590 \textit{ASCA} data yielded $\log L_{2-10\mathrm{keV}} =
40.78$ \citep{bassani_three-dimensional_1999} and later {\it XMM-Newton} $\log
L_{2-10\mathrm{keV}} = 39.70$ \citep{shu_xmm-newton_2010}. This led to a severe over-estimation of the MIR flux and thus an under-estimation of the VISIR exposure times for
these objects.


\section{Observations and data reduction}\label{sec:obs}
For each target a set of two filters was chosen, such that the band-pass of one contained only continuum emission around 12\,$\mu$m while the other contained the polycyclic aromatic hydrocarbon (PAH)
emission feature at 11.3\,$\mu$m. 
As all objects are at very low redshifts ($\langle z \rangle = 0.006$), the NeIIref1 ($\lambda_\mathrm{c}= 12.27\,\mu\mathrm{m}$, half-band width $= 0.18\,\mu\mathrm{m}$) and PAH2 ($\lambda_\mathrm{c}= 11.25\,\mu\mathrm{m}$, half-band width $=
0.59\,\mu\mathrm{m}$) filters were used in all cases. 

The VLT/VISIR observations were carried out in service mode between April and
September 2009 (ESO period P83). Most observations fulfilled the required
observing conditions (airmass $\le 1.3$, optical seeing $\le 0\farcs8$, MIR
photometric sensitivity within 20\% of expected values). An exception is IC
1459, which was observed in conditions that resulted in a much worse sensitivity.
Note that NGC 676 was observed again in P86 with longer exposure times and the
more sensitive PAH2ref2 filter ($\lambda_\mathrm{c}=
11.88\,\mu\mathrm{m}$, half-band width $= 0.37\,\mu\mathrm{m}$). These data will be used here instead of the NeIIref1 data from P83.

Standard chopping and nodding  observing mode with a chop throw of 8\arcsec\, on the small field of view ($19\farcs2 \times 19\farcs2$; 0\farcs075/pixel) was performed. 
Therefore, all four beams fell on the detector -- for science targets the 2 positive on top of each other (parallel mode), ensuring the highest possible signal-to-noise (S/N) for this observing mode. 
After consulting  \textit{Spitzer}/IRAC 8\,$\mu$m images (or 2MASS if the former was not available) the chopping throw and angle for each individual object were chosen such that overlap of the nucleus with non-nuclear emission regions was avoided. 
The flux level of the diffuse host emission at a distance of 8\arcsec drops to less than 10\% compared to the nucleus in 80\% of the cases. 
Thus, possible contamination due to chopping induced overlaps is estimated to be $< 10\%$. 
The most severe exception is NGC 7590, which shows very complex extended MIR morphologies in IRAC. 
Thus, it was observed with the generic nodding template using offset of 30\arcsec, but it remained undetected with VISIR. 
For each science target a photometric standard star was observed  within 2 hours beforehand or
afterwards. The observational data are summarized in
Table~\ref{tab:obs}.

The observations and MIR properties of the archival objects are described in detail
in the literature (\cite{reunanen_vlt_2010} for NGC 1386 and NGC 4486; G+09 for
ESO 005-G004; \cite{horst_mid_2008} for NGC 4303, NGC 4472, NGC 4698, and NGC
5128; and \cite{horst_small_2006} for NGC 4579), except NGC 1052 (programme ID:
382.A-0604, PI: Treister). For all those objects the available raw data were
re-reduced in exactly the same way as described below.

Data reductions were performed using the ESO delivered pipeline. Afterwards,
2-dimensional Gaussians were fitted to all beams of science and calibration
target under the assumption that all targets appear point-like. This will be
justified in Sect.~\ref{sec:MIR}. By subtracting the constant term from the fit
it is possible to extract the unresolved flux from the local background. Thus,
this method is not heavily affected by the background as long as it is locally
smooth. Due to the chosen chop throw, emission structures extended over
more than 4\arcsec, e.g.
the diffuse emission of the host galaxy or weak extended star formation
regions, overlap and are at least partially subtracted. In
particular for the MIR-faint galaxies targeted here, detecting extended sources with
VISIR is difficult. Additional complications arise due to fast local sky background variations of the atmosphere in the MIR and various instrument artifacts that are not understood (see VISIR manual
Sect. 4.9). 

The absolute photometric calibration was obtained by observing standard stars from \cite{cohen_spectral_1999} and computing  conversion factors from the integrated intensity of the beams. 
Its precision is limited by the uncertainties in the fluxes of the standard stars ($\le 10\,\%$). 
In addition, a statistical error can be derived by comparing the
Gaussian fits of the single beams for each target, and is given in Table~\ref{tab:obs}. 
Note that comparison with the aperture method used in \cite{horst_mid_2008} yields consistent results within the uncertainties.

For non-detected nuclei the following procedure was performed: A 2D Gaussian
with the FWHM of the standard star and an amplitude of $2 \sigma_\mathrm{BG}$ of the local
background (50x50 pixel) in the image center was integrated resulting in the
flux upper limit. This method was tested and verified by simulating similar
beams on a large number of artificial and real VISIR images. In all cases the
artificial beam was clearly visible and thus these calculated values represent
upper limits with a confidence of $\ge 99\,\%$. In addition, they
are in accordance with expectations based on the exposure time and the
sensitivity measured from the standard star during the observations.

\begin{table*}
\begin{minipage}[t]{\textwidth}
\caption{Observational parameters and reduction results for the P83 sample.}

\centering 
\label{tab:obs}      
\renewcommand{\footnoterule}{}  
\begin{tabular}{l c c c c c c c c c}        
\hline\hline                 
Object  & Obs. date & Filter& $\lambda_\mathrm{rest}$\footnote{: central filter wavelength in the restframe with half-band width}& Exp. time\footnote{: on-source exposure time excluding any overheads} & \multicolumn{2}{c}{FWHM [\arcsec]} & Flux\\    
            &   (YY-MM-DD) &  & $\mu$m &[s] & STD\footnote{: FWHM of the photometric standard star} & Obj. & [mJy] \\
\hline                        
IC 1459 	&	09-06-05	&	PAH2	&	11.18	$\pm$	0.59	&	362	&	0.29	&	\dots	&		$<$	20	\\
\smallskip	&		&	NeIIref1	&	12.20	$\pm$	0.18	&	1443	&	0.32	&	\dots	&		$<$	30	\\
NGC 676	&	10-12-22	&	PAH2	&	11.19	$\pm$	0.59	&	1448	&	0.32	&	\dots	&		$<$	9	\\
\smallskip	&		&	PAH2ref2	&	11.82	$\pm$	0.37	&	1448	&	0.33	&	\dots	&		$<$	6	\\
NGC 1097	&	09-07-14	&	PAH2	&	11.20	$\pm$	0.59	&	997	&	0.39	&	0.49	&	60	$\pm$	10	\\
\smallskip	&	06-08-07	&	NeIIref1	&	12.22	$\pm$	0.18	&	2524	&	0.32	&	0.40	&	27	$\pm$	4	\\
NGC 1404	&	09-07-15	&	PAH2	&	11.18	$\pm$	0.59	&	544	&	0.57	&	\dots	&		$<$	35	\\
\smallskip	&	09-09-06	&	NeIIref1	&	12.19	$\pm$	0.18	&	1992	&	0.38	&	\dots	&		$<$	18	\\
NGC 1566	&	09-09-06	&	PAH2	&	11.19	$\pm$	0.59	&	362	&	0.37	&	0.46	&	101	$\pm$	12	\\
\smallskip	&		&	NeIIref1	&	12.21	$\pm$	0.18	&	1443	&	0.33	&	0.42	&	95	$\pm$	13	\\
NGC 1667	&	09-09-06	&	PAH2	&	11.08	$\pm$	0.58	&	362	&	0.37	&	\dots	&	$<$ 16	\\
\smallskip	&		&	NeIIref1	&	12.09	$\pm$	0.18	&	1803	&	0.38	&	\dots	&	$<$ 15		\\
NGC 3125	&	09-06-17	&	PAH2	&	11.21	$\pm$	0.58	&	902	&	0.30	&	0.72	&	45	$\pm$	15	\\
\smallskip	&	09-06-16	&	NeIIref1	&	12.22	$\pm$	0.18	&	1992	&	0.32	&	0.74	&	65	$\pm$	13	\\
NGC 3312	&	09-07-15	&	PAH2	&	11.14	$\pm$	0.59	&	725	&	0.38	&	\dots	&		$<$	24	\\
\smallskip	&	\dots	&	NeIIref1	&	12.15	$\pm$	0.18	&	\dots	&	\dots	&	\dots	&			\dots	\\
NGC 4235	&	09-06-07	&	PAH2	&	11.16	$\pm$	0.59	&	362	&	0.29	&	0.38	&	38	$\pm$	6	\\
\smallskip	&		&	NeIIref1	&	12.17	$\pm$	0.18	&	1443	&	0.33	&	0.37	&	33	$\pm$	4	\\
NGC 4261	&	09-06-16	&	PAH2	&	11.17	$\pm$	0.59	&	544	&	0.30	&	0.39	&	12	$\pm$	2	\\
\smallskip	&		&	NeIIref1	&	12.18	$\pm$	0.18	&	1992	&	0.36	&	0.48	&	15	$\pm$	2	\\
NGC 4594\,(M 104)	&	09-06-17	&	PAH2	&	11.21	$\pm$	0.59	&	362	&	0.30	&	\dots	&		$<$	11	\\
\smallskip	&		&	NeIIref1	&	12.23	$\pm$	0.18	&	902	&	0.33	&	\dots	&		$<$	17	\\
NGC 4941	&	09-05-11	&	PAH2	&	11.21	$\pm$	0.59	&	725	&	0.36	&	0.33	&	68	$\pm$	5	\\
\smallskip	&	06-04-19	&	NeIIref1	&	12.22	$\pm$	0.18	&	1620	&	0.35	&	0.35	&	75	$\pm$	4	\\
NGC 5363	&	09-05-10	&	PAH2	&	11.21	$\pm$	0.59	&	544	&	0.33	&	\dots	&		$<$	11	\\
\smallskip	&		&	NeIIref1	&	12.22	$\pm$	0.18	&	1443	&	0.34	&	\dots	&		$<$	18	\\
NGC 5813	&	09-05-10	&	PAH2	&	11.18	$\pm$	0.59	&	181	&	0.33	&	\dots	&		$<$	21	\\
\smallskip	&		&	NeIIref1	&	12.19	$\pm$	0.18	&	181	&	0.34	&	\dots	&		$<$	51	\\
NGC 7213	&	09-05-17	&	PAH2	&	11.18	$\pm$	0.59	&	181	&	0.28	&	0.30	&	232	$\pm$	5	\\
\smallskip	&		&	NeIIref1	&	12.20	$\pm$	0.18	&	181	&	0.31	&	0.32	&	228	$\pm$	15	\\
NGC 7590	&	09-06-04	&	PAH2	&	11.19	$\pm$	0.59	&	362	&	0.32	&	\dots	&		$<$	16	\\
\smallskip	&		&	NeIIref1	&	12.21	$\pm$	0.18	&	902	&	0.34	&	\dots	&		$<$	28	\\
NGC 7626	&	09-06-04	&	PAH2	&	11.12	$\pm$	0.58	&	181	&	0.32	&	\dots	&		$<$	24	\\
\smallskip	&		&	NeIIref1	&	12.13	$\pm$	0.18	&	721	&	0.34	&	\dots	&		$<$	26	\\
NGC 7743	&	09-07-05	&	PAH2	&	11.19	$\pm$	0.59	&	362	&	0.29	&	\dots	&		$<$	11	\\
	&		&	NeIIref1	&	12.20	$\pm$	0.18	&	1443	&	0.32	&	\dots	&		$<$	13	\\

\hline                                   
\end{tabular}
\end{minipage}
\end{table*}

\section{Results}\label{sec:res}

\subsection{MIR morphology and photometry}\label{sec:MIR}
\begin{figure}
      \centering
   \includegraphics[angle=0,width=7cm]{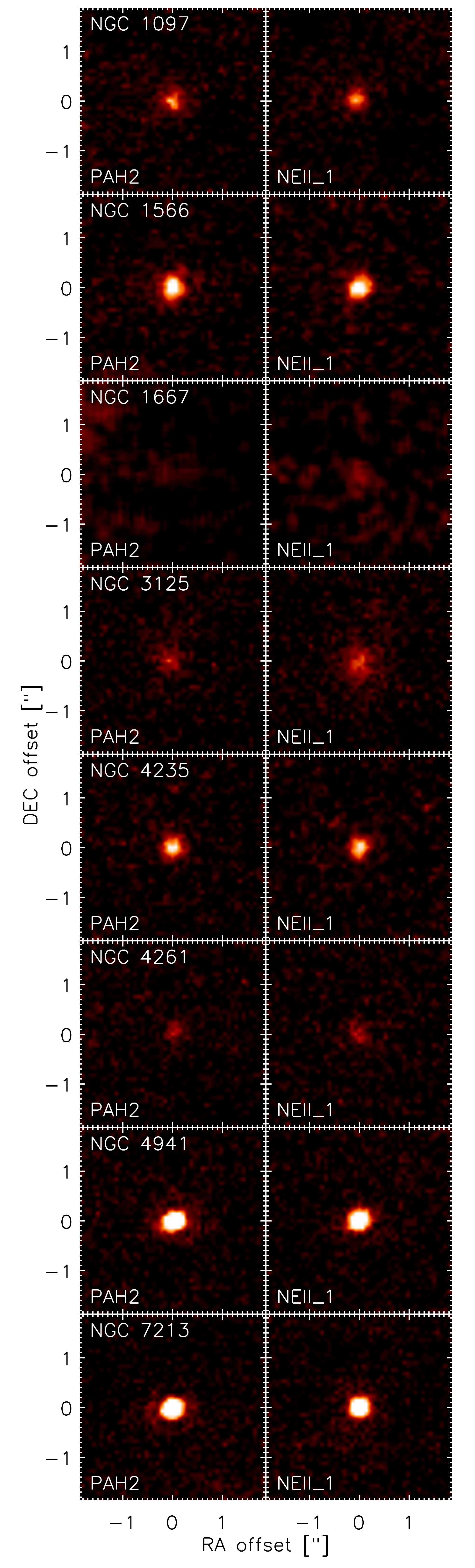}
      \caption{
               VISIR images of the P83 LLAGN in the filters PAH2 and NeIIref1. Each row shows one object sorted by right ascension. White black corresponds to the mean background value ($\langle$BG$\rangle$) and white color corresponds to $\langle$BG$\rangle + 10 \sigma_{\mathrm{BG}}$ with a linear scaling. The images of NGC 1667 were smoothed to increase the visibility of this low-S/N detection.
              }
         \label{fig:im}
   \end{figure}

Of the 17 observed LLAGN, 7 were detected, all in both filters. 
The other 10 and NGC 1404 were not detected in any filter. 
Corresponding linearly scaled images in the PAH2 and NeIIref1 filters are presented in Fig.~\ref{fig:im}. 
These images are created by co-adding sub-images (50 x 50 pixels) of the negative beams (with changed sign) to the sub-image of the central double beam. 
In general, all objects show a point-like central source with a detection significance of  $\sim 5\sigma_{\mathrm{BG}}$ in most cases. 
Here, $\sigma_{\mathrm{BG}}$ signifies the standard variation of the local background ($4\arcsec$). 
The only exception is NGC 3125 which seems to be extended in NeIIref1 and also in PAH2, but still the emission is acceptably fitted  by a Gaussian. 
On the other hand, the detection is at a low $\sigma$-level ($\sim 3$), which further complicates a reliable measurement of its extension.
The corresponding FWHM and fluxes as well as upper limits for the non-detections can be found in Table~\ref{tab:obs}. 
For NGC 1667, individual beams were detected only at a 2 $\sigma_{\mathrm{BG}}$-level, but all 3 beams are located at the expected positions with respect to the chopping
parameters. 
Thus, the detection significance rises to $3.7\sigma_{\mathrm{BG}}$ for the smoothed combined image (as displayed in Fig.~\ref{fig:im}). 
However, the measured fluxes of NGC 1667 will formally be treated as upper limits throughout this work.
NGC 676 was not detected but a foreground star is superposed $5\arcsec$ south of the galaxy center.
The latter was detected in both filters ($5\sigma_{\mathrm{BG}}$, $8$\,mJy in PAH2ref2) constraining the flux of the AGN to 6\,mJy in PAh2ref2 (3$\sigma_{\mathrm{BG}}$ upper limit).
For NGC 4941 and NGC 7213 the Airy rings are evident in the PAH2 images indicating that the diffraction limit was reached during the exposure in those cases. 

Within the pointing accuracy of VISIR all point-source positions are consistent with the ones derived for the galaxy nuclei based on other observations (e.g. 2MASS). 
Thus, it is inferred that the observed point-sources are in fact the MIR imprint of the  LLAGN. 
In addition, no other emission sources of any nature are detected within $8\arcsec$ around the nucleus. 
Furthermore, the science targets show a $\sim 20\%$ larger FWHM compared to the standards.
This fact is related to the much longer exposure times of the science observations which lead to a beam widening due to two different processes: MIR seeing variations on minute time-scales, and instabilities in the VISIR PSF due to instrumental effects as described in \cite{horst_mid-infrared_2009}. 
Both effects can lead to a significant variation of the FWHM during long integration
times. 
In principle, this can be quantified and corrected by comparing and discarding the worst chopping raw frames. 
However, in the case of the LLAGN, this is not possible, as they are too faint to be seen in the individual raw frames.
Therefore, comparisons between standard and science observations are of limited use even if both were performed under similar atmospheric conditions. 

In summary, there is no obvious evidence for extended emission (except NGC 3125), non-nuclear discrete sources or elongation in any LLAGN, but all the observed MIR emission likely originates in a region unresolved by VLT/VISIR. 


\begin{table*}
\begin{minipage}[t]{\textwidth}
\caption{Luminosities and other properties for the whole LLAGN sample.
}

\centering 
\label{tab:lum}      
\renewcommand{\footnoterule}{}  
\begin{tabular}{l c c c c c c c c c c}        
\hline\hline    
Object & AGN type& $D$   & Ref. & $r_0$  &  $\log N_\mathrm{H} $   &  $\log L_{2-10\,\mathrm{keV}}$ & $\log \lambda L_{12\,\mu\mathrm{m}}$ & $\log M_\mathrm{BH}$ & Ref.& $\log \lambda_\mathrm{Edd}$\\
       &         & [Mpc] &     & [pc]    &  $\mathrm{cm}^{-2}$    & [erg/s]                      & [erg/s]                               & [$M_\odot$]          &     &        \\
\hline
NGC 676	&	Sy 2.0	&	19.5	$\pm$	8.7	&	18	&	33	&	$	>	$	24.3	&	40.79	$\pm$	0.60	&		$<$	40.84	&	7.56	&	10	&	$<$	-4.0	\\
NGC 1052	&	L	&	18.0	$\pm$	2.4	&	12	&	31	&	$		$	23.1	&	41.18	$\pm$	0.30	&	42.13	$\pm$	0.07	&	8.19	&	10	&		-3.8	\\
NGC 1097	&	L	&	19.1	$\pm$	4.2	&	19	&	32	&	$		$	20.4	&	40.88	$\pm$	0.15	&	41.23	$\pm$	0.06	&	8.08	&	13	&		-4.2	\\
NGC 1386	&	Sy 2.0	&	18.3	$\pm$	1.4	&	14	&	31	&	$	>	$	24.3	&	42.10	$\pm$	0.60	&	42.26	$\pm$	0.01	&	7.78	&	11	&		-2.9	\\
NGC 1404	&	16LG	&	18.3	$\pm$	1.4	&	14	&	31	&	$	<	$	22.0	&	39.92	$\pm$	0.30	&		$<$	41.25	&	8.50	&	11	&	$<$	-5.2	\\
NGC 1566	&	Sy 1.5	&	14.3	$\pm$	5.9	&	1	&	24	&	$	<	$	22.0	&	41.62	$\pm$	0.30	&	41.76	$\pm$	0.06	&	7.01	&	11	&		-2.6	\\
NGC 1667	&	Sy 2.0	&	62.6	$\pm$	4.3	&	16	&	106	&	$	>	$	24.3	&	42.33	$\pm$	0.60	&	$<$ 42.25	&	7.81	&	11	&		$<$-2.8	\\
ESO 005-G004	&	Sy 2.0	&	22.4	$\pm$	10.0	&	18	&	38	&	$		$	24.0	&	41.82	$\pm$	0.30	&	42.00	$\pm$	0.08	&	7.89	&	20	&		-3.2	\\
NGC 3125	&	L	&	19.8	$\pm$	1.4	&	16	&	34	&	$		$	21.7	&	40.14	$\pm$	0.30	&	41.87	$\pm$	0.08	&	5.77	&	5	&		-2.0	\\
NGC 3312	&	L	&	44.7	$\pm$	3.1	&	16	&	76	&	$	<	$	22.0	&	41.10	$\pm$	0.60	&		$<$	42.15	&	8.28	&	11	&	$<$	-3.9	\\
NGC 4235	&	Sy 1.2	&	35.2	$\pm$	15.7	&	18	&	60	&	$		$	21.2	&	41.61	$\pm$	0.30	&	42.08	$\pm$	0.05	&	7.60	&	10	&		-3.0	\\
NGC 4261	&	L	&	29.4	$\pm$	2.7	&	12	&	50	&	$		$	23.2	&	41.00	$\pm$	0.30	&	41.58	$\pm$	0.05	&	8.72	&	6	&		-4.7	\\
NGC 4303\,(M 61)	&	Sy 2.0	&	15.2	$\pm$	3.1	&	18	&	26	&	$	<	$	22.0	&	39.04	$\pm$	0.30	&	40.72	$\pm$	0.23	&	6.52	&	10	&		-3.9	\\
NGC 4472\,(M 49)	&	Sy 2.0	&	17.1	$\pm$	0.6	&	15	&	29	&	$	<	$	22.0	&	39.00	$\pm$	0.30	&		$<$	41.05	&	8.81	&	10	&	$<$	-6.0	\\
NGC 4486\,(M 87)	&	L	&	17.1	$\pm$	0.6	&	15	&	29	&	$		$	22.6	&	40.73	$\pm$	0.30	&	41.26	$\pm$	0.05	&	9.82	&	9	&		-6.1	\\
NGC 4579\,(M 58)	&	L	&	16.8	$\pm$	3.4	&	18	&	29	&	$		$	21.7	&	41.18	$\pm$	0.15	&	41.70	$\pm$	0.06	&	7.77	&	10	&		-3.6	\\
NGC 4594\,(M 104)	&	Sy 1.9	&	9.1	$\pm$	0.8	&	12	&	15	&	$		$	21.3	&	39.99	$\pm$	0.30	&		$<$	40.61	&	8.46	&	10	&	$<$	-5.4	\\
NGC 4698	&	Sy 2.0	&	23.8	$\pm$	6.4	&	17	&	40	&	$	>	$	24.3	&	40.82	$\pm$	0.60	&		$<$	41.30	&	7.57	&	10	&	$<$	-3.8	\\
NGC 4941	&	Sy 2.0	&	19.8	$\pm$	1.4	&	16	&	34	&	$		$	23.7	&	41.43	$\pm$	0.30	&	41.94	$\pm$	0.02	&	6.91	&	11	&		-2.5	\\
NGC 5128\,(Cen A)	&	Sy 2.0	&	3.4	$\pm$	0.2	&	7	&	6	&	$		$	23.0	&	41.69	$\pm$	0.15	&	41.71	$\pm$	0.01	&	7.74	&	4	&		-3.3	\\
NGC 5363	&	L	&	19.4	$\pm$	1.4	&	16	&	33	&	$	>	$	24.3	&	41.44	$\pm$	0.60	&		$<$	41.30	&	8.36	&	10	&	$<$	-4.2	\\
NGC 5813	&	L	&	32.1	$\pm$	2.8	&	2	&	54	&	$		$	21.1	&	38.77	$\pm$	0.30	&		$<$	42.19	&	8.53	&	10	&	$<$	-5.3	\\
NGC 7213	&	L	&	21.2	$\pm$	1.5	&	16	&	36	&	$		$	20.3	&	42.09	$\pm$	0.15	&	42.48	$\pm$	0.03	&	7.74	&	11	&		-2.7	\\
IC 1459 	&	Sy 2.0	&	30.3	$\pm$	4.2	&	2	&	51	&	$		$	22.1	&	40.61	$\pm$	0.30	&		$<$	41.91	&	9.41	&	3	&	$<$	-5.4	\\
NGC 7590	&	Sy 2.0	&	27.4	$\pm$	6.0	&	19	&	47	&	$	<	$	22.0	&	39.77	$\pm$	0.60	&		$<$	41.79	&	6.71	&	8	&	$<$	-3.2	\\
NGC 7626	&	L	&	41.9	$\pm$	2.9	&	16	&	71	&	$	<	$	22.0	&	39.56	$\pm$	0.30	&		$<$	42.13	&	8.78	&	10	&	$<$	-5.2	\\
NGC 7743	&	Sy 2.0	&	19.2	$\pm$	1.6	&	12	&	33	&	$	>	$	24.3	&	41.27	$\pm$	0.60	&		$<$	41.15	&	6.64	&	10	&	$<$	-2.7	\\

\hline                                   
\end{tabular}
\end{minipage}
\newline 
{\it -- Notes:} References on distances and black hole masses.
(1) average of \cite{tully_nearby_1988}, \cite{willick_homogeneous_1997} and $D_L$ from NED;
(2) \cite{blakeslee_synthesis_2001};
(3) \cite{cappellari_counterrotating_2002}; 
(4) \cite{cappellari_mass_2009};
(5) \cite{dudik_chandra_2005};
(6) \cite{ferrarese_discovery_1996}; 
(7) \cite{ferrarese_discovery_2007}; 
(8) \cite{garcia-rissmann_atlas_2005}; 
(9) \cite{gebhardt_black-hole_2011}; 
(10) \cite{ho_search_2009}; 
(11) Hyperleda database \citep{paturel_hyperleda._2003}; 
(12) \cite{jensen_measuring_2003}; 
(13) \cite{lewis_black_2006}; 
(14) \cite{madore_hubble_1999}; 
(15) \cite{mei_acs_2007};
(16) $D_L$ from NED; 
(17) \cite{springob_sfi++._2007}; 
(18) \cite{tully_nearby_1988}; 
(19) \cite{willick_homogeneous_1997}; 
(29) \cite{winter_optical_2010}. 
$r_0$: spatial scale corresponding to the theoretical diffraction limit for a 8.2\,m VLT telescope at 12\,$\mu$m. 
$\log \lambda = \log (L_\mathrm{Bol} / L_\mathrm{Edd})$ is the Eddington ratio.
\end{table*}

\subsection{MIR-X-ray relation}
In the following, the P83 data are combined with the archival data described above (Sect.~\ref{sec:obs}) to include all LLAGN detected by VLT/VISIR so far. 
In order to compare the positions of the LLAGN in the luminosity plane with respect to the MIR-X-ray correlation the monochromatic 12\,$\mu$m continuum luminosities $\lambda L_{\lambda}(12\mu\mathrm{m})$ (hereafter $L_\mathrm{MIR}$) are computed by using the NeIIref1 fluxes.
For this, a flat MIR SED (in $\lambda L_\lambda$) around 12\,$\mu$m was assumed and no $K$-correction was applied. 
Due to the low redshifts of the sample the latter effect is negligible.
As already described in Sect.~\ref{sec:xray}, the intrinsic hard X-ray luminosities $L_{2-10\,\mathrm{keV}}$  (hereafter $L_\mathrm{X}$)  are inferred from the literature.
The corresponding luminosity values can be found in Table~\ref{tab:lum} for the distance set c).

The MIR-X-ray correlation strength is measured by using the Spearman Rank correlation coefficient, $\rho$, and
the corresponding null hypothesis probability, $p$ (\texttt{r\_correlate} in IDL).
To quantify the correlation, linear regression in logarithmic space is
performed. Due to the presence of a variety of systematic uncertainties (see discussion in G+09) and a significant number of upper
limits it is useful to compare different fitting methods. 
Here, the Bayesian based \texttt{linmix\_err} \citep{kelly_aspects_2007} and the canonical \texttt{fitexy} \citep{press_numerical_1992}, will be used. 
Both algorithms include treatment of errors on both axes. 
In addition, \texttt{linmix\_err} can handle upper limits and intrinsic scatter as well, which makes it the superior algorithm for this investigation. 
It uses a Markov chain Monte Carlo method to draw random parameter sets from the probability distributions given by the measured data. 
The maxima of the resulting distributions of parameter draws represent the best-fit values \citep[details and comparison tests can be found in][]{kelly_aspects_2007}. 
Here, typically $10^5$ such random draws will be used. 
Apart from an estimate of the intrinsic scatter $\sigma_\mathrm{int}$, \texttt{linmix\_err} also returns a linear correlation strength, comparable to the Spearman Rank. 
On the other hand, the Spearman Rank and the \texttt{fitexy} will also be stated for easier comparison with previous work. 
Note that the X-ray luminosity will always be treated as the independent variable. 
This is physically motivated, as it originates in the innermost region around the black hole directly tracing the accretion, while the MIR emission is originating likely by UV heated dust further out. 

\subsubsection{MIR-X-ray relation for LLAGN}\label{sec:LLAGN}
The absorption-corrected X-ray versus MIR luminosities for all LLAGN are displayed with filled symbols (or arrows in case of upper limits) in Fig.~\ref{fig:L+G} together with the AGN from G+09.
To indicate the sensitivity limits due to the observational layout (see Sect.~\ref{sec:sample}), hatched areas are displayed in the plot as well. 
Objects in those areas could not be detected at a significant level in the P83 VISIR programme.
Their borders are determined for each X-ray luminosity by simulating the choice of an exposure time for an hypothetical point-source at a representative distance.
This, in turn, results in a MIR luminosity for which the detection significance would drop below $3\sigma_{\mathrm{BG}}$. 
These areas represent only a rough guideline and are not necessarily valid for every individual object, especially in those cases where the exposure times were too short due to outdated X-ray luminosities (see Sect.~\ref{sec:xray}). 
The latter explains why some upper limits (e.g. NGC 5813 and NGC 7626) are so far above the sensitivity limits.

Although the LLAGN sample does not extend over a large luminosity range, the MIR and intrinsic X-ray luminosities are highly correlated. 
However, 2 outliers become evident by exhibiting offsets $> 1$\,dex towards high MIR  (or low X-ray) luminosities: NGC 3125 and NGC 4303 (solid grey symbols in Fig.~\ref{fig:L+G}). 
Both will be excluded form the correlation and regression analysis which will be justified in Sect.~\ref{sec:out}.
For the luminosities of the other LLAGN, the Spearman Rank coefficient varies between 0.78 and 0.88 (corresponding to a null-hypothesis probability $ -2.54 \ge \log p \ge -3.78$ ), depending on the distance set used.
The LLAGN luminosities have been fitted with the two algorithms described above, and the \texttt{fitexy} fit to all detected LLAGN is displayed as dashed line in Fig.~\ref{fig:L+G}.
The results of both fitting algorithms are in general agreement (compare Table~\ref{tab:cor}) while \texttt{linmix\_err} gives systematically flatter slopes and larger uncertainties for the parameters for these samples. 
Note that the fits including upper limits are likely biased due to due to the outdated X-ray luminosities leading to many very high upper limits.
Furthermore, the \texttt{fitexy} fit parameters for the different distance sets do not all agree within the calculated uncertainties.
Thus, in general, the slope of the MIR-X-ray correlation for LLAGN is only constrained to be between $\sim$0.9 and 1.1.
Note that although the Compton-thickness of some objects (e.g. NGC 7590 and NGC 7743) remains uncertain, they do not significantly affect the results. This is because they are not detected with VISIR and their upper limits do not constrain the results with either high (absorbed) or low (unabsorbed) intrinsic X-ray luminosities.

\subsubsection{Comparison to brighter AGN}\label{sec:comp}

The main goal of this investigation is to determine whether LLAGN deviate from the MIR-X-ray correlation found for brighter local AGN (G+09). 
The latter are plotted in Fig.~\ref{fig:L+G} with empty symbols, while ``well-resolved'' AGN are in addition marked with a central black-filled circle (see G+09 for details). 
These ``well-resolved'' local Seyferts are probably the least contaminated by non-AGN emission, and will be used for comparison with the LLAGN. 
Their sample properties are repeated in Table~\ref{tab:cor} and the corresponding power-law fit is as well displayed in Fig.~\ref{fig:L+G} as a dotted line.
In general, the LLAGN are offset by $\sim0.3$\,dex from the best fit of the brighter AGN towards higher MIR (or lower X-ray) luminosities. 
In addition, the slopes of the LLAGN fits are all flatter (in the X-ray vs MIR plane), and in particular for the $D_L$ and TF distance sets.
However, the uncertainties of the LLAGN fit parameters are so large in all other cases, that the 1-$\sigma$ confidence intervals overlap with the brighter AGN fits.
Thus, there is no large systematic offset between the LLAGN and AGN. 
Furthermore, the 6 detected LINERs constrain any possible difference of this class to the Seyfert class to be $< 1$\,dex in the MIR-X-ray luminosity plane.

\subsubsection{MIR-X-ray correlation for all AGN}\label{sec:all}
The previous section showed that the LLAGN do not differ significantly from brighter AGN.
Thus, it makes sense to investigate the properties of the combined AGN sample (LLAGN + all AGN from G+09).
The corresponding correlation and regression properties are displayed in the bottom part of Table~\ref{tab:cor}.
The correlation strength measured by the Spearman rank coefficient significantly increases from $\rho=0.88$ to 0.92, and becomes insensitive to the choice of the distance set or even inclusion of the outliers. 
The same is true also for the fitexy parameters, which are consistent within the uncertainties for all distance sets and sample combinations. 
In particular, the fit with distance set c) is indistinguishable from the one for all AGN in G+09 (displayed as solid line in Fig.~\ref{fig:L+G}).
Comparing the fitting algorithms \texttt{fitexy} and \texttt{linmix\_err}, the later gives again systematically lower values for the fitted parameters, but consistent with the former.

The MIR-X-ray correlation is also present in the flux-flux space (see Fig.~\ref{fig:flux}). The intrinsic 2-10 keV flux is calculated with the luminosities and distances given in Table~\ref{tab:lum}. Here, the formal correlation strength is weaker ($\rho = 0.72 ; \log p = -8.0$), due to the decreased range in orders of magnitude compared to the luminosity space.
The slope flux correlation is very similar to that of the luminosity correlation for the combined sample (using \texttt{fitexy}):
\begin{equation}
 \log \left( \frac{F_\mathrm{MIR}}{\mathrm{mJy}}\right)   = ( 14.35 \pm 0.82) + ( 1.12 \pm 0.08 )  \log \left( \frac{F_\mathrm{X}}{\mathrm{erg}\,\mathrm{s}^{-1}\mathrm{cm}^{-2}}\right),
\end{equation} 
whereas again no $K$-correction is applied. 
The strong flux correlation and its similar slope prove that the luminosity correlation is not a consequence of a selection bias related to distance. 

Summarizing, there is no significant evidence for a deviation of the LLAGN as a group.
Instead, they are consistent with the MIR-X-ray luminosity correlation found in previous work for brighter AGN (e.g. G+09). 
Thus, the main result of this section is, that the MIR-X-ray correlation extends unchanged down to luminosities of $ \lesssim 10^{41}$\,erg/s with a small scatter.

\subsubsection{Comparison to starburst galaxies}
\begin{figure*}

      \sidecaption  
 \includegraphics[angle=0,width=12cm]{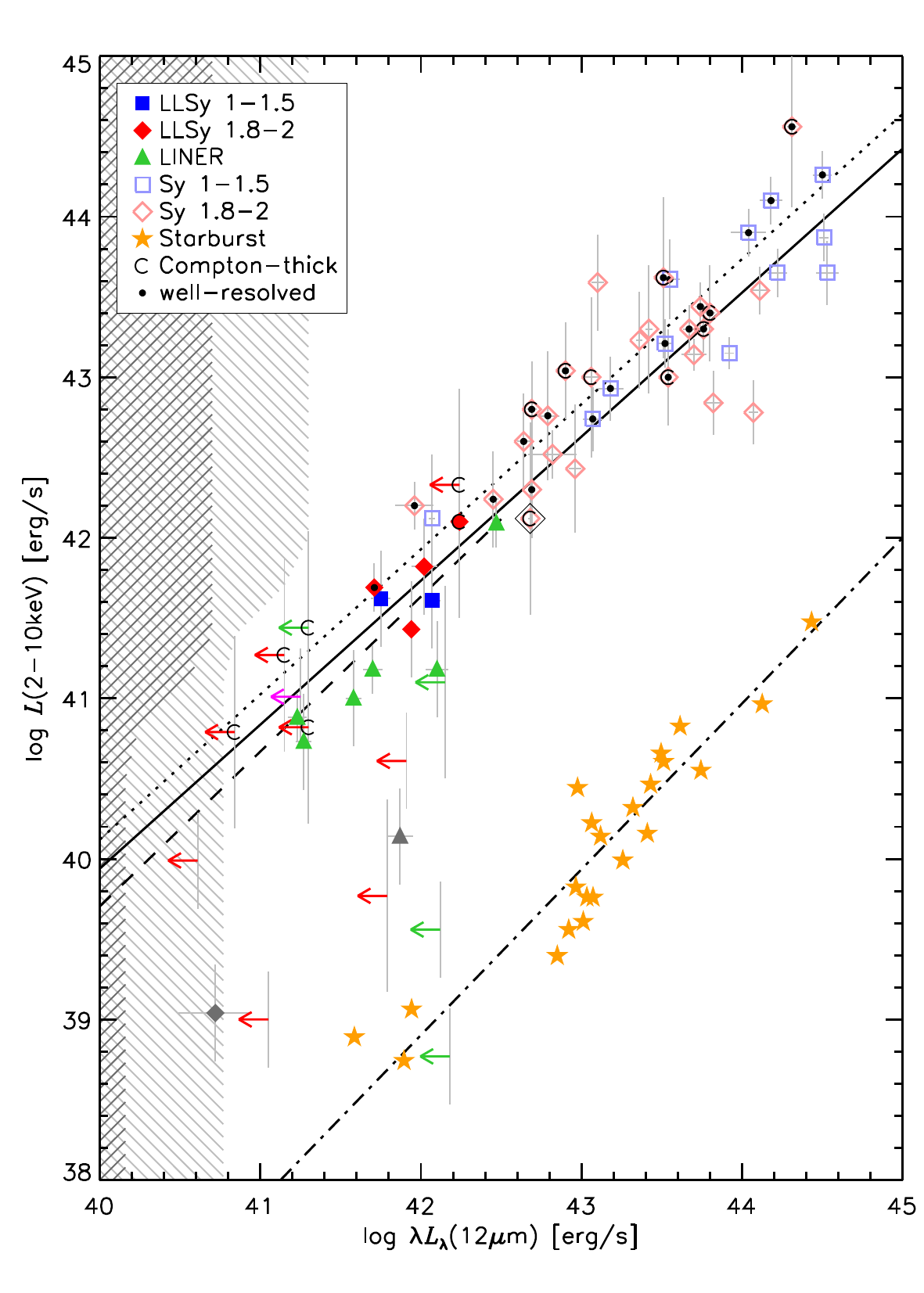}
      \caption{  
               Absorption-corrected hard X-ray luminosities vs. the nuclear MIR
               luminosities for all LLAGN (filled symbols) and the AGN from G+09 (empty symbols);
               blue squares: type 1 Seyferts (type 1.5 or smaller),
               red diamonds: type 2 Seyferts; green triangles: LINERs;
               magenta arrow: NGC 1404 (NELG);
               Circinus is plotted for comparison only (framed in black);
               objects marked with a C: Compton thick AGNs (X-ray $N_{\rm H}>1.5 \cdot 10^{24}$\,cm$^{-2}$);
               objects highlighted with central black-filled circles: ``well-resolved'' AGN from G+09;
               NGC 3125 and NGC 4303 are displayed in grey; 
               filled orange stars: starburst galaxies (total luminosities) from \cite{ ranalli_2-10_2003};
               solid line: power-law fit to all LLAGN and AGN displayed in color
               dashed line: power-law fit to all LLAGN (except NGC 3125 and NGC 4303);
               dotted line: power-law fit to ``well-resolved'' AGN from G+09;
               dot-dashed line: power-law fit to all SB.
               The hatched areas indicate regions below the detection limit of S/N=3 for the performed VISIR NeIIref1 imaging: light grey: objects at 20\,Mpc; dark grey objects at 10\,Mpc; see text for details. 
              }
         \label{fig:L+G}
   \end{figure*}

\begin{figure}
      \centering
   \includegraphics[angle=0,width=0.8\columnwidth]{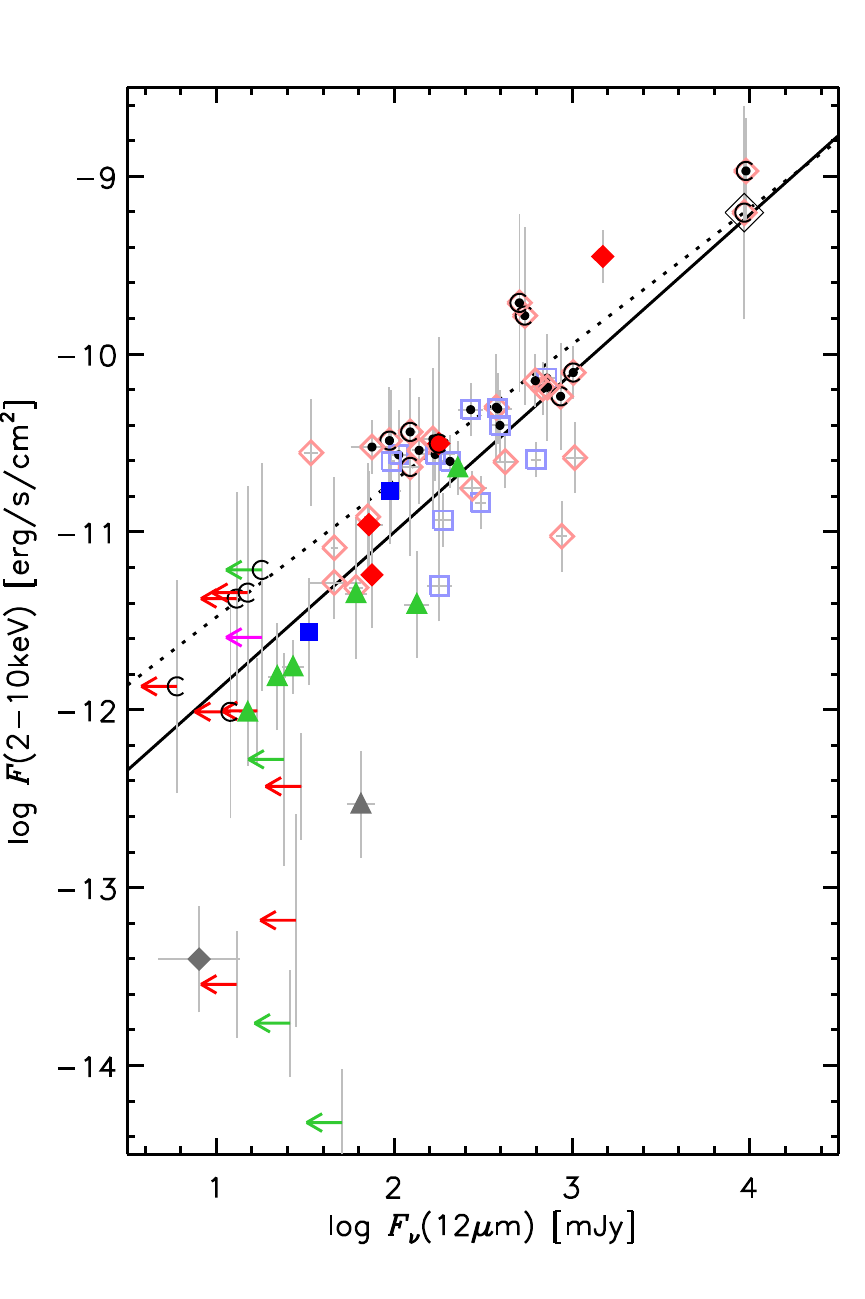}
      \caption{
                Correlation of MIR and absorption-corrected hard
                X-ray fluxes for the extended LLAGN sample and AGN from
                G+09. 
                Description is similar to Fig.~\ref{fig:L+G}.
              }
         \label{fig:flux}
   \end{figure}

It is interesting to compare the MIR-X-ray properties of the AGN to the ones of typical
starburst galaxies (SB) in which the MIR emission is dominated by star formation. 
Here, the galaxy sample of \cite{ranalli_2-10_2003} is used, consisting of 22 typical nearby starburst galaxies, all possessing measured 2-10\,keV X-ray luminosities presented in the same work. 
The hard X-ray emission in these objects originates mainly in X-ray binaries and shock heated diffuse gas. 
As the star formation in these galaxies is occurring on global and not (only) nuclear scales, large aperture photometry is used for the comparison (\textit{IRAS} fluxes at 12\,$\mu$m). 
The SBs are displayed as orange stars in Fig.~\ref{fig:L+G} while the corresponding power-law fit is denoted as the dot-dashed line with parameters listed in Table~\ref{tab:cor}. 
No luminosity uncertainties are given in \cite{ranalli_2-10_2003}, thus the bisector algorithm was used.   
The SB form a strong correlation ($\rho = 0.92$; $\log p = -9.0$) with a high luminosity ratio ($\overline{r} = 3.1$) and a small scatter ($\sigma_r=0.25$) in agreement with \cite{krabbe_n-band_2001}.
They form a population clearly distinct from the AGN.
Interestingly, the slope ($0.97 \pm 0.07$) of the power-law fit is comparable to the one of the AGN, although the X-ray emission originates from completely different mechanisms.
On the other hand, the normalization of the MIR-X-ray correlation for the SB galaxies differs by $\sim 2.6$ orders of magnitude, towards higher MIR (or lower X-ray) emission.  

\begin{table*}
\begin{minipage}[t]{\textwidth}
\caption{Correlation properties between $\log L_\mathrm{MIR}$ and $\log
L_\mathrm{X}$ to various sample populations. 
}

\centering 
\label{tab:cor}      
\renewcommand{\footnoterule}{}  
\begin{tabular}{l c c c c c c c c c}        
\hline\hline    
	Sample 	&	Method	&	$N$	&	$\rho$	& $\log p$ &	$\sigma_\mathrm{int}$
&	$a$			&	$b$		 &	$\overline{r}$	&	$\sigma_r$	\\ 
\hline										
	G+09(well-resolved) 	&	fitexy	&	22	&	0.93			&	-9.52	&	\dots			&	0.19	$\pm$	0.05	&	1.11	$\pm$	0.07	&	0.15	&	0.23	\\
	G+09(well-resolved) 	&	linmix\_err	&	22	&	0.99	$\pm$	0.02	&	\dots	&	0.12	$\pm$	0.15	&	0.18	$\pm$	0.06	&	1.08	$\pm$	0.09	&	0.15	&	0.23	\\
	LLAGN(detected)	&	fitexy	&	12	&	0.78			&	-2.54	&	\dots			&	0.42	$\pm$	0.27	&	1.04	$\pm$	0.17	&	0.40	&	0.25	\\
	LLAGN(detected)	&	linmix\_err	&	12	&	0.92	$\pm$	0.16	&	\dots	&	0.20	$\pm$	0.25	&	0.16	$\pm$	0.44	&	0.85	$\pm$	0.28	&	0.40	&	0.25	\\
	LLAGN(all)	&	linmix\_err	&	25	&	0.98	$\pm$	0.05	&	\dots	&	0.25	$\pm$	0.36	&	0.44	$\pm$	0.46	&	1.09	$\pm$	0.27	&	0.40	&	0.25	\\
	LLAGN(detected; $D_L$)	&	fitexy	&	12	&	0.88			&	-3.78	&	\dots			&	0.06	$\pm$	0.12	&	0.74	$\pm$	0.08	&	0.42	&	0.25	\\
	LLAGN(detected; TF)	&	fitexy	&	12	&	0.84			&	-3.14	&	\dots			&	0.15	$\pm$	0.17	&	0.84	$\pm$	0.10	&	0.42	&	0.25	\\
	LINER(detected)	&	fitexy	&	6	&	0.93			&	-2.12	&	\dots			&	0.49	$\pm$	0.32	&	1.01	$\pm$	0.17	&	0.55	&	0.20	\\
	LINER(detected)	&	linmix\_err	&	6	&	0.89	$\pm$	0.34	&	\dots	&	0.34	$\pm$	2.76	&	0.39	$\pm$	1.49	&	0.94	$\pm$	0.81	&	0.55	&	0.20	\\
\smallskip
	LINER(all)	&	linmix\_err	&	10	&	0.94	$\pm$	0.24	&	\dots	&	0.38	$\pm$	5.39	&	0.39	$\pm$	2.00	&	0.96	$\pm$	1.10	&	0.55	&	0.20	\\

	G+09(all)	&	fitexy	&	42	&	0.88			&	-14.01	&	\dots			&	0.41	$\pm$	0.03	&	1.12	$\pm$	0.04	&	0.30	&	0.36	\\
	G+09(all)	&	linmix\_err	&	42	&	0.95	$\pm$	0.03	&	\dots	&	0.29	$\pm$	0.19	&	0.35	$\pm$	0.06	&	1.00	$\pm$	0.08	&	0.30	&	0.36	\\
	LLAGN(detected) \& G+09(all)	&	fitexy	&	48	&	0.92			&	-19.02	&	\dots			&	0.41	$\pm$	0.03	&	1.12	$\pm$	0.04	&	0.31	&	0.35	\\
	LLAGN(detected) \& G+09(all)	&	linmix\_err	&	48	&	0.96	$\pm$	0.02	&	\dots	&	0.26	$\pm$	0.17	&	0.35	$\pm$	0.06	&	1.02	$\pm$	0.06	&	0.31	&	0.35	\\
	LLAGN(all) \& G+09(all)	&	linmix\_err	&	61	&	0.98	$\pm$	0.01	&	\dots	&	0.27	$\pm$	0.17	&	0.34	$\pm$	0.06	&	1.06	$\pm$	0.06	&	0.31	&	0.35	\\
        SB                      & bisector & 22 & 0.92 & -8.97 & \dots & 2.96 $\pm$ 0.20 & 0.97 $\pm$ 0.07 & 3.06 & 0.25 \\

\hline

\end{tabular}

\end{minipage}
\newline 

        {\it -- Notes:} Methods: see text for a description; for 'linmix\_err' the given values are medians of the parameters;
         all uncertainties are given as standard deviations;
        $N$: number of objects used for the analysis; 
        $\rho$: linear correlation coefficient (for 'fitexy': Spearman Rank); 
        $p$: null-hypothesis probability; 
        $\sigma_\mathrm{int}$: intrinsic scatter (estimated by 'linmix\_err');
        $a,b$: fitting parameters of $\log L_\mathrm{MIR}-43 = a + b (\log
L_\mathrm{X} -43)$; 
        $\overline{r}$: average of the luminosity ratio $r = \log
(L_\mathrm{MIR}/L_\mathrm{X})$ with $\sigma_r$ its standard deviation.

\end{table*}

\section{Discussion}
 \label{sec:dis}

\subsection{Comparison with \textit{IRAS}}\label{sec:IRAS}
\begin{figure}
      \centering
   \includegraphics[angle=0,width=0.8\columnwidth]{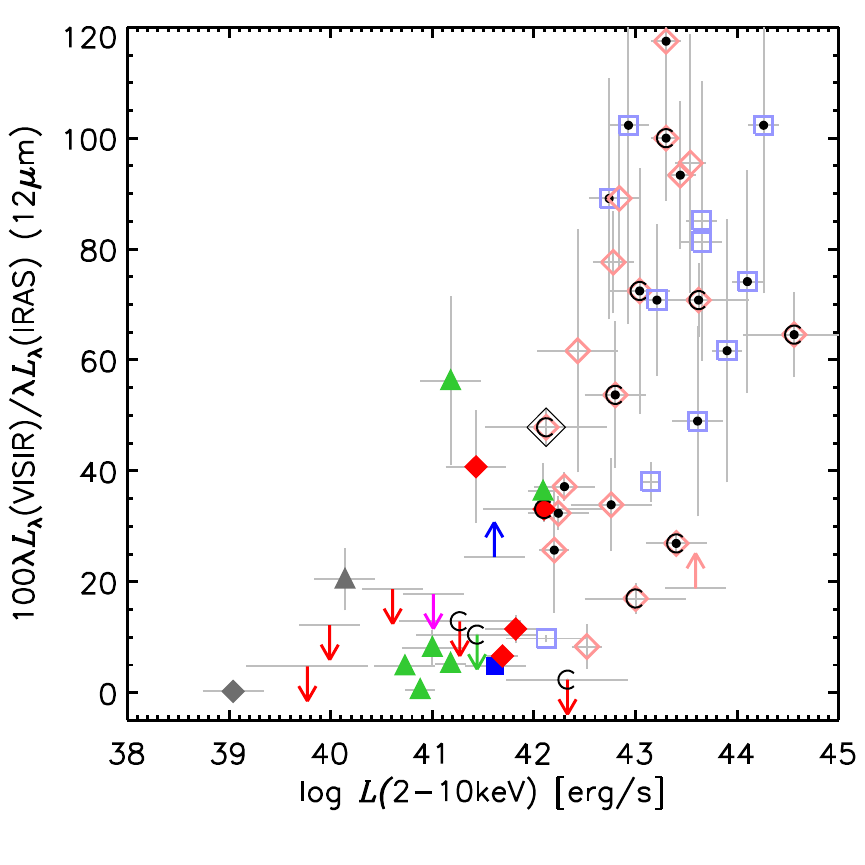}
      \caption{
                Fraction of total 12\,$\mu$m luminosity that is emitted by the nucleus. It is measured by the ratio of VISIR to \textit{IRAS} luminosities versus the  absorption-corrected hard
                X-ray luminosity for the extended LLAGN sample and AGN from
                G+09. 
                Description is similar to Fig.~\ref{fig:L+G}, except the dashed line, which is a power-law fit (linmix\_err) to the combined AGN sample.
              }
         \label{fig:IRAS}
   \end{figure}

The obtained high spatial resolution data give the opportunity to study the AGN contribution to the total 12\,$\mu$m emission of nearby galaxies. 
The latter can be measured by using the large-aperture \textit{IRAS} photometry (FWHM $\sim30\,\arcsec$), taken either from \cite{sanders_iras_2003} or the NED database.
Fig.~\ref{fig:IRAS} shows the ratio of the 12\,$\mu$m luminosities measured on nuclear scale with VISIR and global scale with \textit{IRAS} over the absorption corrected X-ray luminosities for all LLAGN and the AGN from G+09. 
In general, the nuclear fraction of the total MIR luminosity is increasing with increasing X-ray luminosity but with large scatter of a factor $\sim 2.5$.
The 2 objects showing the lowest ratio, NGC 1097 and NGC 4303, are both dusty spiral galaxies, the former dominated by its well-known circum-nuclear starburst ring.
Thus, as expected, the total MIR emission of the galaxies is dominated by non-nuclear emission and the AGN contribution dominates only at X-ray luminosities $\gtrsim 10^{43.6}$\,erg/s, while  at $L_\mathrm{X} \sim 10^{42}$\,erg/s the host contribution to the MIR is already 3 to 10 times larger than the one from the nucleus. 
This was already found by \cite{vasudevan_power_2010} in the logarithmic space, where a strong correlation ($\rho = 0.70, \log p = -6.4 $) is present, and  emphasizes once more the need for high spatial resolution to isolate the AGN from surrounding processes.

\subsection{Comparison to \textit{Spitzer}/IRS}\label{sec:IRS}
Most of the LLAGN have also been observed with \textit{Spitzer}/IRS, which enables a more detailed comparison of the innermost 4\arcsec.
In particular, low spectral resolution N-band spectra will be used here. 
Post-BCD spectra are available in the archive, created by a point-source profile extraction. 
They are well-calibrated (uncertainty $\lesssim 10 \%$) and are sufficient for a rough comparison with the nuclear properties, which are the focus of this paper.
It is important to note, however, that these IRS spectra can significantly deviate from other studies, due to 2 possible reasons: in the case of extended objects the extraction profile under-weights the extended emission, or the flux might be significantly lower due to bigger extraction areas (e.g. $20\arcsec$ in \cite{gallimore_infrared_2010}).
Nevertheless, for the objects in common, NGC 1097 and NGC 1566, the post-BCD spectra are in general agreement with those published by \cite{diamond-stanic_effect_2010}, where very small extraction areas have been used.

\begin{figure*}
      \centering
   \includegraphics[angle=0,width=16cm]{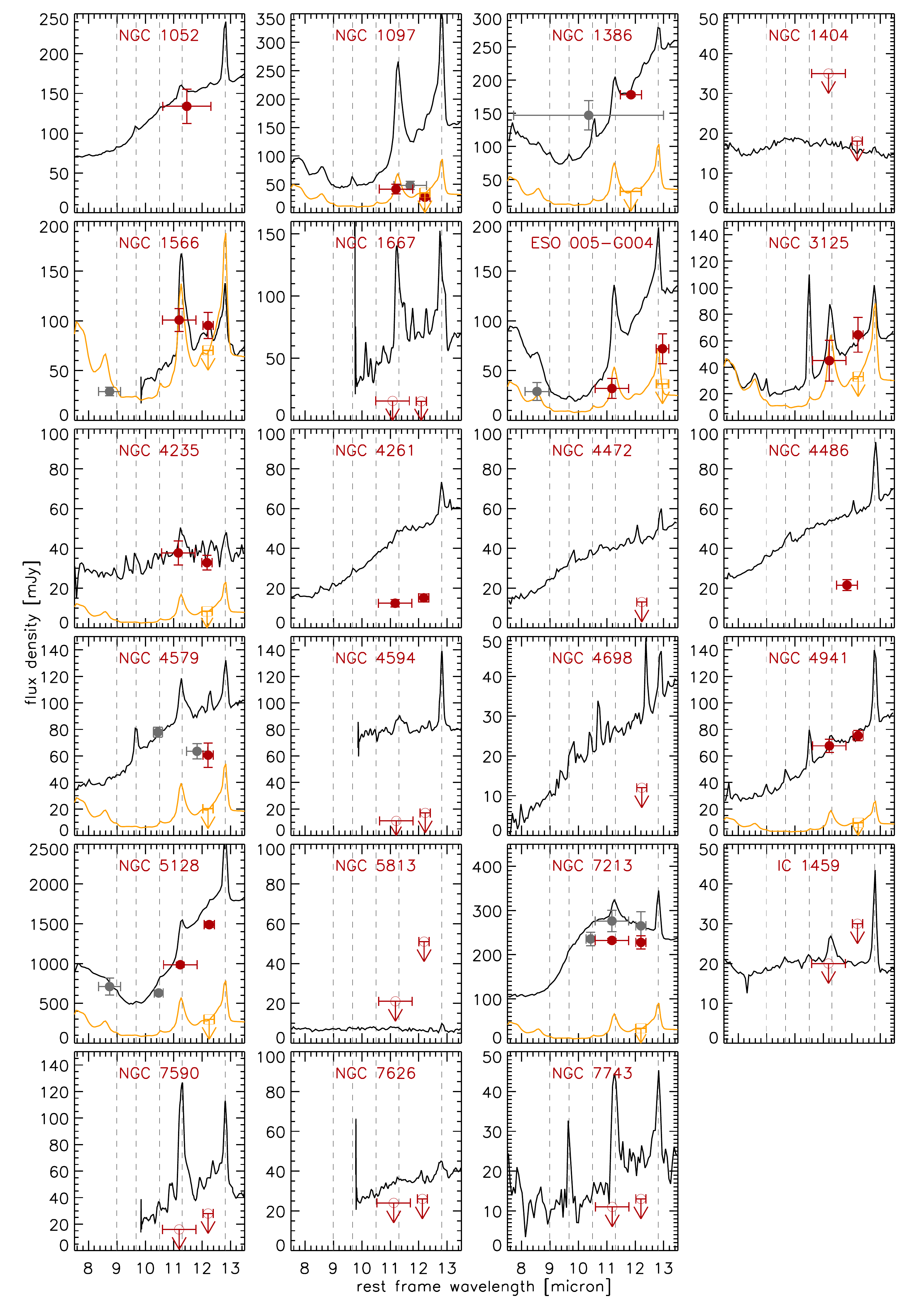}
      \caption{
               Comparison of VISIR photometry (red symbols: P83 data , gray
symbols: re-reduced data from the archive) and {\it Spitzer} IRS
spectra (black line). In addition, photometric measurements in the narrow-N filter for NGC 1097 \citep{mason_dust_2007}, and the Si2 filter
for NGC 1386 and NGC 5128 and in the N filter for NGC 1386 are also plotted in
gray for comparison \citep[]{ramos_almeida_infrared_2009}.
Horizontal error bars correspond to the filter pass band.
The IRS spectra of NGC 1566, NGC 4594, NGC 1667, NGC 7590, and NGC 7626 are smoothed for better visibility. 
Common emission lines are indicated by the dotted lines. For determination of the rest frame wavelengths redshifts from the NED database are used.
In addition, scaled starburst template SED (displayed as orange line) are shown as well as the derived upper limit for the starburst contribution to the continuum filter measurement (orange symbols), see Sect.~\ref{sec:PAH} for details.
              }
         \label{fig:SEDs}
   \end{figure*}

The data are displayed in Fig.~\ref{fig:SEDs} while, the flux uncertainties of the IRS spectra ($\lesssim 10\%$) are not displayed to maintain clarity. 
Common MIR emission lines are indicated by vertical dashed lines, from left to right:
[ArIII] (8.99\,$\mu$m), H$_2$ (9.67\,$\mu$m), [SIV] (10.5\,$\mu$m), PAH
(11.3\,$\mu$m) and [NeII] (12.81\,$\mu$m). 
For NGC 676, NGC 3312, and NGC 5363 no IRS spectra are available in the archive. 
Furthermore, for NGC 1667, NGC 4594, NGC 7590 and NGC 7626, the IRS spectra are in high resolution mode with a low S/N, although in some cases a strong PAH feature (11.3\,$\mu$m) is still evident.
In addition, archival VISIR imaging data along with the photometric measurements of
\cite{ramos_almeida_infrared_2009} an \citep{mason_dust_2007} are shown for comparison (gray data points). 
They show good agreement with the new measurements within the uncertainties. 
Only NGC 7213 shows different fluxes ($\gtrsim 10\%$) in both NeIIref1 and PAH2 over a time of $\sim$3 years.
The reason for this is unknown but might indicate variability. 

For comparison with the VISIR photometry ($F_\nu^\mathrm{VISIR}$), the IRS spectra are convolved with the normalized filter transfer functions of PAH2 and NeIIref1. 
The resulting fluxes, $F_\nu^\mathrm{IRS}$(PAH2) and $F_\nu^\mathrm{IRS}$(NeIIref1), are displayed in Table~\ref{tab:IRS} for all LLAGN detected in both filters.
There are many cases, where Spitzer displays a much higher 12\,$\mu$m continuum flux than VISIR at a higher spatial resolution, especially for NGC 1097, NGC 4261 (also NGC 4486). 
This is also true for the upper limits derived for NGC 1667, NGC 4594 and NGC 7590 and implies that large amounts of hot dust are distributed in extended diffuse regions around the nucleus. 
Due to their faintness, the latter are probably resolved-out in VISIR and thus not evident. 
Unfortunately, the S/N of the images is not high enough to constrain these emission regions well, e.g., in
the case of NGC 4261, the missing flux of $\sim 35$\,mJy extended over a region of $0.75\arcsec$ diameter would become undetectable in VISIR.  
However, for the other objects (NGC 1566, NGC 3125, NGC 4235 and NGC 4941) the VISIR and IRS continuum fluxes are very similar, indicating that in these cases no additional significant MIR source is surrounding the nucleus at arcsec scale. 

The imaging in the second filter setting (PAH2) was performed to constrain the strength of the PAH emission feature at 11.3\,$\mu$m, which is commonly used as a star formation indicator \citep[e.g.][]{diamond-stanic_relationship_2011}.
In particular, for NGC 1566 the convolved IRS flux ($F_\nu^\mathrm{IRS}(\mathrm{PAH2}) = 103 \pm 10$\,mJy) is very similar to VISIR PAH2 flux ($F_\nu^\mathrm{VISIR}(\mathrm{PAH2}) = 100.9 \pm 11.5$\,mJy).
This indicates the presence of PAH emission inside the inner $0.4\arcsec$. 
On the other hand, the weak PAH emission features present in the IRS spectra of NGC 4235 and NGC 4941 cannot be verified in the VISIR data, as the measurement uncertainties of are too large.
For NGC 1097,   the VISIR fluxes are both much lower than in the IRS spectrum -- the latter showing very strong PAH emission.
However, the PAH2 flux in VISIR is much higher than in NeIIref1 implying that the nucleus of NGC 1097 also emits large amounts of PAH emission.
Nevertheless in general, the lower continuum fluxes and absence of PAH emission in VISIR indicates, that this feature typical for star-formation usually does not originate in the very nucleus of LLAGN, but in the circum-nuclear surroundings (at scales larger than 100\,pc). 
This was already noticed by \cite{hoenig_dusty_2010-1} for Seyfert galaxies. 
These diffuse emission regions are likely very extended, so that they are resolved-out on most VISIR images.

However, to further constrain any star formation contamination of the VISIR data, in the next section, the most extreme possible case -- namely that all the PAH emission (seen in the IRS spectra) is originating in the unresolved nucleus -- will be examined.

\begin{table*}

\begin{minipage}[t]{\textwidth}
\caption{Comparison between VISIR and \textit{Spitzer}/IRS flux densities.}
\label{tab:IRS}
\centering       
\begin{tabular}{l c c c c c c c}        
\hline\hline   
Object  & $F_\nu^\mathrm{VISIR}$(PAH2) & $F_\nu^\mathrm{IRS}$(PAH2) & $R$(PAH2)  & $F_\nu^\mathrm{VISIR}$(NeIIref1) &
$F_\nu^\mathrm{IRS}$(NeIIref1) &$R$(NeIIref1)\\    
            & [mJy]  & [mJy] &   & [mJy] & [mJy] & \\
\hline 

NGC 1097	&	60.4	$\pm$	10.2	&	168.4	&	2.8 &	27.3	$\pm$	4.2	&	158.2	&	5.8
\\ 
NGC 1566	&	100.9	$\pm$	11.5	&	103.2	&	1.0 &	95.4	$\pm$	13	&	94.0	&	1.0
\\
NGC 3125	&	45	$\pm$	15.4	&	60.2	&	1.3 &	64.5	$\pm$	13.1	&	59.2	&	0.9
\\
NGC 4235	&	37.7	$\pm$	6	&	41.1	&	1.1 &	32.8	$\pm$	3.7	&	39.6	&	1.2
\\
NGC 4261	&	11.8	$\pm$	2	&	46.0	&	3.9 &	15.1	$\pm$	2	&	51.2	&	3.4
\\
NGC 4941	&	67.6	$\pm$	4.9	&	68.0	&	1.0 &	75.3	$\pm$	3.9	&	76.5	&	1.0
\\
NGC 5128	&	984.8	$\pm$	32.1	&	1322.9	&	1.3 &	1488.7	$\pm$	35.1	&	1767.7	&	1.2
\\
NGC 7213	&	232.2	$\pm$	4.5	&	297.5	&	1.3 &	227.7	$\pm$	15	&	263.2	&	1.2
\\

\hline
\end{tabular}

\end{minipage}
\newline 

        {\it -- Notes:} The given IRS fluxes are extracted from the BCD spectra
convolved with the normalized VISIR filter transfer functions of PAH2 and NeIIref1; the PAH2
fluxes contain not only the PAH feature at 11.3\,$\mu$m but also the underlying continuum;
$R$(PAH2) and $R$(NeIIref1) give the flux ratios
($F_\nu^\mathrm{IRS}/F_\nu^\mathrm{VISIR}$) for both filters.

\end{table*}

\subsection{Constraining the nuclear star formation contamination}\label{sec:PAH}
One of the main concerns of this investigation is whether the measured 12\,$\mu$m continuum is contaminated by non-AGN emission, e.g. circum-nuclear star formation.
Such contamination would increasingly affect the measurements at lower luminosities, i.e. as in the case of the LLAGN. 
It could in fact explain the MIR-excess that some LLAGN show.
To better constrain the individual star formation contamination, the LLAGN data can be compared to a typical star formation SED. 
Such template SED can be created by using, e.g., the IRS spectra of typical nearby starburst galaxies from \cite{brandl_mid-infrared_2006}. 
These authors collected a sample of 13 SB spectra with highest S/N. 
The latter have been made  available electronically by the \textit{Spitzer}/IRS Atlas project \citep{hernan-caballero_atlas_2011}. 
To create the SB template SED, these 13 SB are normalized by their PAH 11.3\,$\mu$m emission line flux.
This approach is valid because the 12\,$\mu$m continuum  scales with the PAH 11.3\,$\mu$m flux in starbursts as demonstrated by a strong correlation for the whole sample of \cite{brandl_mid-infrared_2006} ($\rho =0.98$ with $\log p = -7.1$): 
 \begin{eqnarray}
 \log \left( \frac{F_\nu(\mathrm{NeIIref1})}{\mathrm{Jy}}\right)   &=&  ( 11.82 \pm 0.91) \cr &+& ( 1.06 \pm 0.08 )  \log \left( \frac{F(\mathrm{PAH}11.3\mu\mathrm{m})}{\mathrm{erg}\,\mathrm{s}^{-1}\mathrm{cm}^{-2}}\right).
\end{eqnarray}
A comparable correlation was recently found by \cite{diamond-stanic_relationship_2011}.
Note that this normalization results in a scatter of approximately 30\% (0.13\,dex) for the 12\,$\mu$m flux in the individual spectra.
The created SB template SED is then scaled for each individual LLAGN by its PAH 11.3\,$\mu$m emission line flux.
Next, the corresponding 12\,$\mu$m flux of the scaled SB is determined by convolution with the continuum filter function (NeIIref1 in most cases).
By this means, it is possible to estimate the maximum contamination by star formation in the measured continuum of the unresolved LLAGN nuclei in VISIR. 

The measured total PAH 11.3\,$\mu$m emission line fluxes $F(\mathrm{PAH} 11.3\mu\mathrm{m})$ are given in Table~\ref{tab:PAH} for all detected LLAGN with evident PAH emission, while the typical measurement error is  $\lesssim 10\%$. 
The scaled starburst template spectra for those LLAGN are plotted in orange in Fig.~\ref{fig:SEDs}.
The corresponding calculated 12\,$\mu$m continuum fluxes $F_\nu^\mathrm{SB}$(12$\mu$m) are indicated by orange open squares and stated in Table\ref{tab:PAH} as well. 
Note that NGC 1052 is not included because it was not observed in a VISIR filter containing only continuum, but instead in the B11.4 filter.
Although this filter includes the PAH emission feature in its band-pass, the latter is very weak in the IRS spectrum of NGC 1052.
Furthermore, for the detected LINERs NGC 4261 and NGC 4486, no significant PAH emission could be detected in the IRS spectra. 
Thus, any significant contribution of star formation to the VISIR data of NGC 1052, NGC 4261 and NGC 4486 is very unlikely.
On the other hand, the IRS spectra of NGC 1097 and NGC 1566 show strong PAH emission and thus their estimated maximum star formation contribution $c^\mathrm{SB}_<$ is very large (100\% and 73\% respectively). 
Note, however, that in the case of NGC 1097 and ESO 005-G004, $F_\nu^\mathrm{SB}$(12$\mu$m) was actually constrained more by the VISIR PAH2 flux, as the latter was lower than the scaled flux of the SB template. 
Nevertheless, in the case of NGC 1097, $F_\nu^\mathrm{SB}$(12$\mu$m) still exceeds the measured 12\,$\mu$m continuum flux in VISIR.
However, the PAH emission feature is weak in all the other cases (NGC 1386, NGC 4235, NGC 4579, NGC 4941, NGC 5128 and NGC 7213), so that the maximal star formation contribution $c^\mathrm{SB}_<$ can be constrained to often much less than 30\%.
In summary, a major contribution by star formation to the nuclear 12\,$\mu$m continuum (at 0.4\arcsec scale) can be excluded for the majority of LLAGN (9 of 12 cases). 
Note that exclusion of the other 3, possibly contaminated, LLAGN does not affect the MIR-X-ray correlation. 

However, it is very unlikely that all the seen PAH emission originates in the unresolved nucleus (as assumed in this section).
Consequently, the real star formation contribution is probably much lower in all cases, as long as the PAH 11.3\,$\mu$m remains a valid tracer of the former.
In particular, it is not verified that the hard AGN radiation not only destroys the PAH molecules but also quenches star formation. 
On the other hand, PAH molecules might survive in the dense clumpy clouds inside the torus itself for significant timescales.
In addition, young massive star clusters emit hard radiation fields that could destroy the PAH molecules \citep[see e.g.][]{snijders_subarcsecond_2006}.
At least, recent findings suggest that the PAH 11.3\,$\mu$m remains on average a good star formation tracer also in circum-nuclear environments of AGN \citep[e.g.][]{diamond-stanic_effect_2010}. 

\begin{table}
\begin{minipage}[t]{\columnwidth}
\caption{Estimation of maximum star formation contribution.}
\label{tab:PAH}
\renewcommand{\footnoterule}{}  
\centering       
\begin{tabular}{l c c c}        
\hline\hline   
Object  & $\log F(\mathrm{PAH} 11.3\mu\mathrm{m})$ & $F_\nu^\mathrm{SB}$(12$\mu$m) & $c^\mathrm{SB}_<$\footnote{: $c^\mathrm{SB}_<$ signifies the maximum star formation contribution (in percent) to the nuclear 12\,$\mu$m continuum flux (at 0.4\arcsec scale).} \\    
            & [erg/s/cm$^2$]  & [mJy] & \% \\
\hline 
       NGC 1097 & -11.95 &  35.2\footnote{: $F_\nu^\mathrm{SB}$(12$\mu$m) of NGC 1097 constrained by the VISIR PAH flux, see text for details.} & 100\\

       NGC 1386 & -12.47 &  32.3 &  18\\
       NGC 1566 & -12.21 &  70.4 &  73\\
   ESO 005-G004 & -12.37 &  36.7\footnote{: $F_\nu^\mathrm{SB}$(12$\mu$m) of ESO 005-G004  constrained by the VISIR PAH2 flux, see text for details.} &  50\\
       NGC 3125 & -12.54 &  32.9 &  51\\
       NGC 4235 & -13.13 &   8.4 &  25\\
       NGC 4579 & -12.76 &  20.2 &  33\\
       NGC 4941 & -13.08 &   9.6 &  12\\
       NGC 5128 & -11.59 & 297.4 &  19\\
       NGC 7213 & -12.53 &  33.3 &  14\\

\hline
\end{tabular}
\end{minipage}
\end{table}

\subsection{Accretion rates}\label{sec:rat}

   \begin{figure*}
      \sidecaption
   \includegraphics[angle=0,width=12cm]{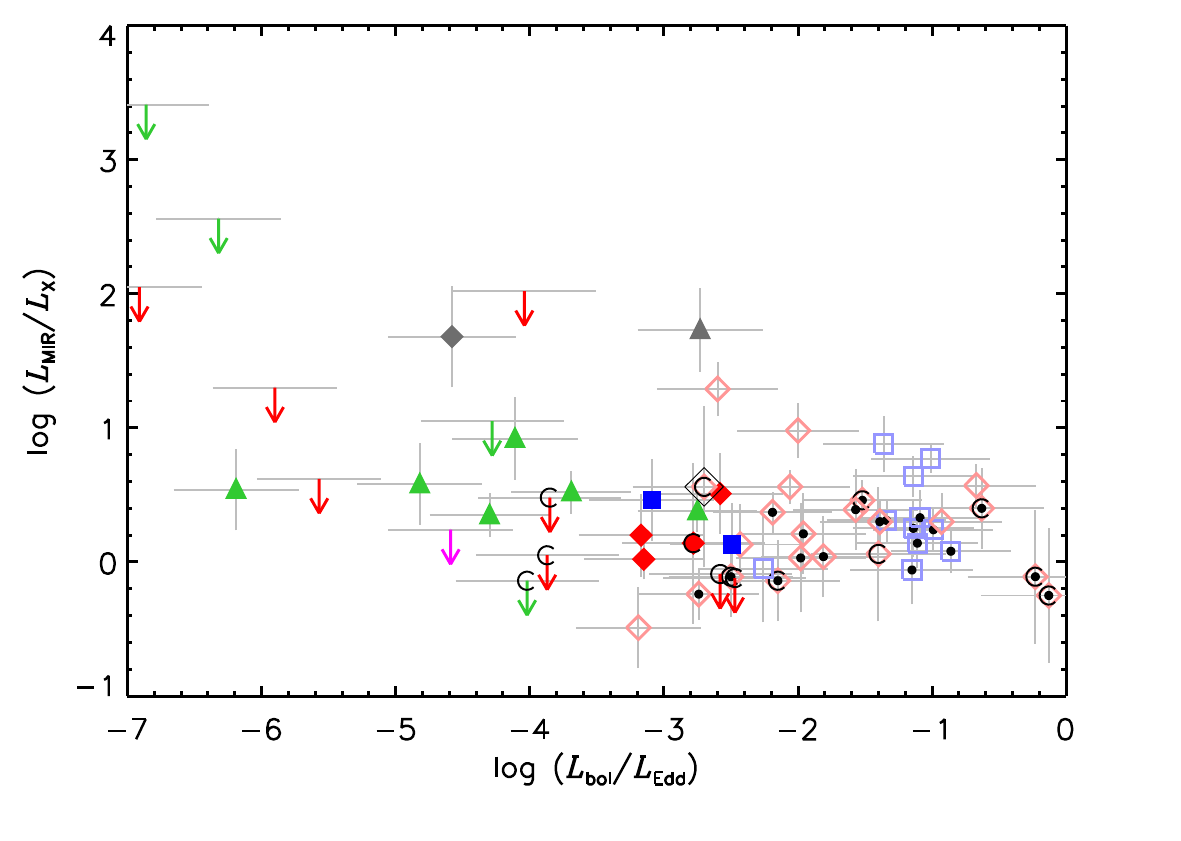}
      \caption{
               12\,$\mu$m to 2-10\,keV luminosity ratio versus Eddington ratio
$\log \lambda_\mathrm{Edd} = \log (L_\mathrm{Bol} / L_\mathrm{Edd})$ for all AGN with black
hole mass estimates available. Symbols are similar to Fig.~\ref{fig:L+G}.
              }
         \label{fig:rate}
   \end{figure*}

Having shown that most LLAGN are not dominated by nuclear star formation in the MIR, the relation between the MIR emission and other AGN properties can be studied, in particular the accretion rate. 
Here, the Eddington ratio will be used as a proxy of the  accretion rate: $\lambda_\mathrm{Edd} = L_\mathrm{Bol} /
L_\mathrm{Edd}$ with $L_\mathrm{Bol}$ the bolometric luminosity and
$L_\mathrm{Edd} = 1.26 \cdot 10^{38} (M_\mathrm{BH} /M_\odot) $\,erg/s being the
Eddington luminosity. 
For estimating $L_\mathrm{Bol}$ the X-ray luminosity is often used
\citep[e.g.,][]{vasudevan_piecing_2007,vasudevan_optical--x-ray_2009} with
$L_\mathrm{Bol} = 10 L_{2-10\,\mathrm{keV}}$ being a reasonable choice, whereas
the exact value of the bolometric correction factor does not affect the main
conclusions of this study. 
The black hole masses used in this analysis were compiled from the literature with preference given to
the most recent determinations, and are listed in Table~\ref{tab:lum} and
Table~\ref{tab:AGN} for the AGN from G+09. In most cases the only available mass
estimate comes from using the empirical $M_\mathrm{BH}-\sigma_*$
relation \citep{gebhardt_relationship_2000,
ferrarese_fundamental_2000,tremaine_slope_2002}, and in those cases the given
reference is for the stellar velocity $\sigma_*$. Furthermore, the newest
determination of the relation \citep{gueltekin_m-_2009} is used, for which the
intrinsic scatter is estimated to be 0.44\,dex. Here, this value will be adopted
as general uncertainty for $M_\mathrm{BH}$.

Fig.~\ref{fig:rate} shows the MIR-X-ray luminosity ratio, 
$ r = \log (L_\mathrm{MIR}/L_\mathrm{X})$, distribution versus the
computed accretion rate.
In general, the observed LLAGN have low accretion rates ( $\lambda_\mathrm{Edd} \le -2.5$), whereas the low-luminosity Seyferts overlap with the brighter AGN. 
On the other hand, the LINERs get clearly separated with lowest accretion rates. 
The latter is believed to be a critical parameter determining the AGN structure. 
As already mentioned, for low accretion rates several changes in the AGN structure are suggested, e.g., for the accretion disk. 
In addition, \cite{falcke_jet_2000} argued that the jet might dominate the AGN emission at low accretion rates and presented a jet model that explains the observed SED of Sgr A*. 
A similar model was used by \cite{markoff_results_2008} to successfully fit the extensive multiwavelength data of M 81, a well-known LLAGN. 
These changes in the AGN structure will also be reflected in the mid-infrared. 
Although neither the accretion disk nor the jet base can be resolved with VLT/VISIR, if any of the two dominates the nuclear MIR emission, evidence should be present in the data.
In particular, \cite{yuan_radio-x-ray_2005} predict, that the jet (if present)
should dominate the X-ray emission at rates below $\lambda_\mathrm{Edd}^\mathrm{crit} \approx
10^{-5}$. 
This is for example the case for NGC 4486, NGC 4594 and IC 1459, while NGC 1052, NGC 4261 and NGC 4579 should be ADAF dominated. 
All these objects are also in the sample of \cite{yuan_revisiting_2009}, who successfully test the above prediction by SED fitting. 
In the latter work no torus model is included and instead the MIR emission is mainly produced by the ADAF. 
Then, the MIR-X-ray luminosity ratio is predicted to be roughly one for the ADAF dominated sources and significantly larger for the jet dominated sources NGC 4486 and NGC 4594. 
But this difference in the ratios is not evident in Fig.~\ref{fig:rate}.

In general, the average luminosity ratio is higher for LLAGN ($\overline{r} = 0.40 \pm 0.25$) compared to the values for the well-resolved ($0.15 \pm 0.23$) and  for all AGN from G+09 ($0.30 \pm 0.36$; see also Table~\ref{tab:cor}). 
In particular, LINERs have a high luminosity ratio ($\overline{r} = 0.55 \pm 0.20$) compared to the type 1 ($\overline{r} = 0.34 \pm 0.29$) and the type 2 Seyferts ($\overline{r} = 0.23 \pm 0.38$).
However, due to the large scatter any differences remain statistically insignificant. 
As expected, NGC 4303 and NGC 3125 are located at peculiar positions in this graph, especially the latter possesses an atypically high accretion rate for a LINER ($\lambda_\mathrm{Edd} = 10^{-2}$). 
In summary, despite the different object types the MIR-X-ray luminosity ratio can be described as constant for all AGN with a scatter of 0.35\,dex over at least 4 (possibly 6) orders of magnitude in accretion rate ($10^{-6} \lesssim \lambda_\mathrm{Edd} \lesssim 1$).

\subsection{The outliers}\label{sec:out}
Two of the objects from the original LLAGN sample exhibit very peculiar positions in the different plots. They are detected only very weakly at $\sim 3\sigma_\mathrm{BG}$ and were excluded from the correlation and regression analysis, because of evidence that their observed fluxes at multiple wavelengths are heavily dominated by non-AGN processes. These two objects are now discussed separately.

\subsubsection{\object{NGC 3125}}
The X-ray analysis of Chandra data published by \cite{dudik_chandra_2005} indicated an AGN in this object, while \cite{zhang_census_2009} reports a point-like X-ray nucleus embedded in soft emission based on the same data. 
On the other hand, this galaxy is known to harbor young and extreme Wolf-Rayet star clusters, two of which have been confirmed by optical and near-infrared observations \citep{hadfield_how_2006}, one directly its galactic nucleus. 
In addition, no broad optical emission lines have been detected and it is optically classified as starburst \citep{kewley_optical_2001}.
Thus, it is unclear, if NGC 3125 really harbors an AGN, or if the X-ray emission originates from X-ray binaries or an ULX inside or near the star cluster. 
The MIR emission may either originate from dust heated by an AGN or by the central star cluster, on which the VISIR imaging was centered. 
The latter is favored by the fact that this object is also the only spatially resolved one in the LLAGN sample. 
Note that the second and fainter star cluster at $\sim 10\,\arcsec$ to the East was not detected. 
Comparison of the VISIR photometry and the IRS spectrum shows good agreement in the continuum indicating that the emission seen by IRS originates from a compact nucleus. 
On the other hand, the strong PAH emission feature in the IRS spectrum seems to be absent in the VISIR  data.
By assuming that the PAH 11.3\,$\mu$m emission is still present in the nucleus the maximum contribution of star formation  $c^\mathrm{SB}_<$ is only $\le 51\%$.
However, complex molecules are probably also destroyed in close proximity of extremely massive young star forming regions. 
Evidence is strong that the Wolf-Rayet star cluster dominates the MIR emission of NGC 3125, being the reason for its peculiar position with respect to the luminosity correlations. 
Hence, the observed MIR flux should be regarded as an upper limit for any dusty torus emission of a putative AGN in NGC 3125.

\subsubsection{\object{NGC 4303} - M 61}
\begin{figure}
      \centering
   \includegraphics[angle=0,width=\columnwidth]{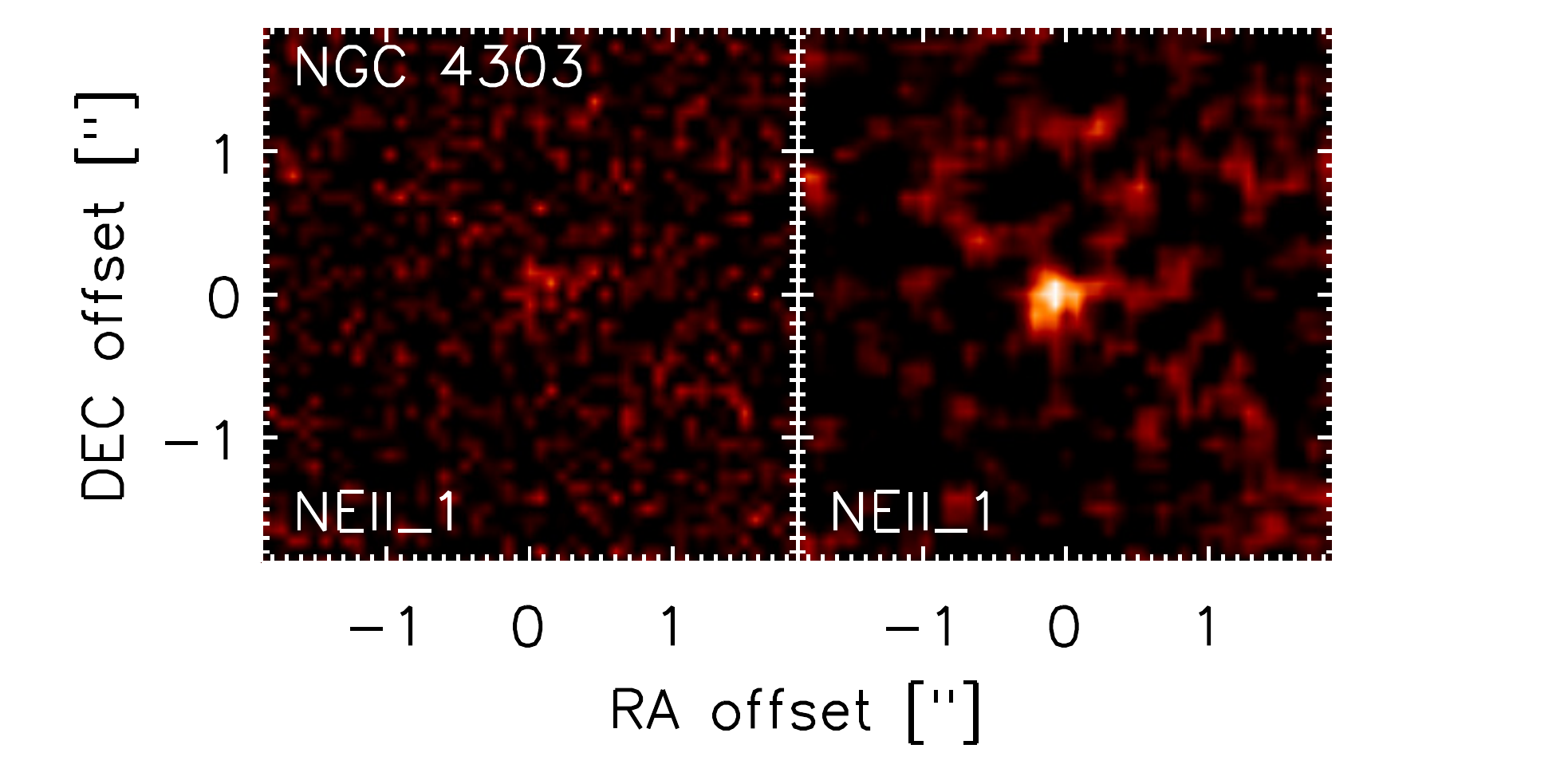}
      \caption{
              VISIR image of NGC 4303 in NeIIref1. Left side: co-added image (50 x 50 pixel); right side: the same image but smoothed before co-adding; white color corresponds to $\langle$BG$\rangle + 5 \sigma_{\mathrm{BG}}$ with a linear scaling
              }
         \label{fig:NGC4303}
   \end{figure}

This object was claimed to be undetected in \cite{horst_mid_2008}. A careful re-analysis of the NeIIref1 image revealed a very weak detection at a low $\sigma\sim 2$ (compare Fig.~\ref{fig:NGC4303}). Similar to NGC 1667, the negative beams were found at the expected positions. Furthermore, the detection significance increases in the smoothed co-added image. Still, the MIR values calculated for this object are unreliable due to the very low S/N. As the flux of this object is very low, it will unfortunately be very difficult to improve the S/N with any further observations.
NGC 4303 was classified as having a HII nucleus by \cite{ho_search_1995}, although the optical line ratios are close to the AGN regime.  Its X-ray luminosity is determined by \cite{tzanavaris_searching_2007} based as well on Chandra data. They conclude that this object is a good AGN candidate even with such a low luminosity and without a detected Fe K$\alpha$ emission line. \cite{jimenez-bailon_nuclear_2003} derive a 0.3\,dex lower value from the same data. Both works report a hard X-ray core surrounded by a softer emission environment possibly associated with a circum-nuclear starburst ring. \cite{colina_detection_2002} find evidence for a young star cluster present in the nucleus. The latter could contribute significantly or even dominate the MIR emission at such low luminosities. This would explain the objects extreme position in the luminosity plane far off the correlation. Unfortunately, there is no IRS spectrum covering the N-band region. 
As for NGC 3125 the nuclear star cluster might easily dominate the observed MIR emission leading to the observed excess. Note that off-nuclear young clusters with comparable MIR luminosities are found for example in NGC 1365 \citep{galliano_extremely_2008} and would be similarly unresolved at the distance of NGC 4303.

\section{Conclusions}\label{sec:con}
VLT/VISIR mid-infrared (N-band) imaging of 17 nearby LLAGN (and one NELG) is presented at a spatial resolution of $\sim 0\farcs35$ in two narrow-band filters. 
For the 7 detections, the images show point-like sources, and no extended emission is found nearby the nucleus. 

Possible sources of this unresolved MIR emission are nuclear star formation, the putative dusty torus, the accretion disk, the jet or the narrow-line region. 
Any significant contribution to the 12\,$\mu$m continuum by the latter is considered unlikely \citep[]{groves_infrared_2006, mor_dusty_2009}. 
The other possibilities are investigated by 3 different methods:

Firstly, comparison to the MIR-X-ray luminosity correlation found for brighter AGN:
Using X-ray luminosities taken from the literature, all detections plus 9 additional LLAGN from the archive are compared to the sample from G+09. 
The MIR-excess (or X-ray deficit) of the LLAGN is not statistically significant.
Instead, the MIR-X-ray correlation from G+09 is also valid for the LLAGN, and the latter extends down to luminosities of the order of of $ < 10^{41}$\,erg/s with a small scatter. 
It is noteworthy that the local starburst galaxies with globally measured MIR and X-ray luminosities show a similar correlation with a comparable slope close to 1 but with $\sim 2.6$ orders of magnitude difference in normalization.
The 2 peculiar outliers, NGC 3125 and NGC 4303, are explained by dominant nuclear star clusters. 

Secondly, comparison to \textit{IRAS} and \textit{Spitzer}/IRS shows that the \textit{IRAS} photometry is dominated by the host galaxy in most cases, with decreasing AGN contribution with decreasing X-ray luminosities. 
The comparison to small-aperture IRS N-band spectra yields much better constraints on the nuclear star formation. 
In particular, the PAH 11.3\,$\mu$m emission line flux is regarded as a star formation indicator.
A typical starburst template SED scaled by the line flux observed in IRS for the individual LLAGN constrains the maximum 12\,$\mu$m continuum contribution of a putative nuclear starburst. 
For 2 objects, NGC 1097 and NGC 1566, strong PAH emission is indicated even at $0.4\arcsec$ scale. 
Star formation might dominate their nuclei in the MIR, but in general the weakness or absence of the PAH feature in 75\% of the LLAGN restricts the nuclear non-AGN contamination to $\lesssim 30\%$, often much less, and thus disfavors star formation as the dominating MIR source in LLAGN. 

Thirdly, for low accretion rates a change in the accretion structure of the AGN is proposed by several works, which should also affect the MIR properties of the LLAGN. 
However, the MIR-X-ray luminosity ratio is independent of the accretion rate, over at least 4 (possibly 6) orders of magnitude down to $\lambda_\mathrm{Edd} \sim 10^{-5}$.

These results do not prove the presence of a dusty torus, but they are consistent with its existence in all observed LLAGN. 
Thus, the unification model might still hold for LLAGN, as already suggested by, e.g., \cite{panessa_x-ray_2006} and \cite{maoz_low-luminosity_2007}. 
Furthermore, \cite{elitzur_disappearance_2009} recently determined a threshold for the disappearance of the torus at very low accretion rates depending on the bolometric luminosity. 
None of the detected LLAGN is below this threshold.
On the other hand, if torus would be absent and the MIR emission would be dominated by the jet or ADAF, then there is a fortuitous agreement in both slope and normalization with the correlation for brighter (torus-dominated) Seyferts.
In any case, the fact that all AGN (Seyferts and LINERs) can be well-described by a single correlation without any strong offsets makes the MIR-X-ray correlation a very useful observational tool for converting between MIR and X-ray powers, irrespective of the nature of the individual AGN.

Increasing the sample size, in particular for the LINERs, would lead to much tighter constraints on the nature of the MIR emission in LLAGN. 
Unfortunately, current MIR facilities either lack the sensitivity or spatial resolution necessary to observe most LLAGN. On the other hand, further investigations using additional multiwavelength data will yield further clues on the nature of the MIR emission and properties of LLAGN.

\begin{acknowledgements}
We would like to thank the anonymous referee for many helpful suggestions that improved this work a lot. In addition, we also thank Sera Markoff for very valuable comments.
This research made use of the NASA/IPAC Extragalactic Database
(NED) which is operated by the Jet Propulsion Laboratory, California Institute
of Technology, under contract with the National Aeronautics and Space
Administration. We acknowledge the usage of the HyperLeda database (http://leda.univ-lyon1.fr).
PG acknowledges a JAXA International Top Young Fellowship. This research is supported in part by a JSPS Grant in Aid number 21740152. 
S.H. acknowledges support by Deutsche Forschungsgemeinschaft (DFG) in the framework of a research fellowship (``Auslandsstipendium'').
DA thanks Michael West for improving suggestions on manuscript. He thanks Amelia Bayo for providing the IRS fluxes and the many motivating comments.
\end{acknowledgements}
\bibliographystyle{aa} 
\bibliography{my_lib_ref.bib} 

\begin{thebibliography}{149}
\expandafter\ifx\csname natexlab\endcsname\relax\def\natexlab#1{#1}\fi

\bibitem[{Akylas \& Georgantopoulos(2009)}]{akylas_xmm-newton_2009}
Akylas, A. \& Georgantopoulos, I. 2009, A\&A, 500, 999

\bibitem[{{Alonso-Herrero} {et~al.}(2011){Alonso-Herrero}, {Ramos Almeida},
  Mason, Asensio~Ramos, Roche, Levenson, Elitzur, Packham,
  Rodr\'iguez~Espinosa, Young, {D\'iaz-Santos}, \&
  {P\'erez-Garc\'ia}}]{alonso-herrero_torus_2011}
{Alonso-Herrero}, A., {Ramos Almeida}, C., Mason, R., {et~al.} 2011, ApJ, 736,
  82

\bibitem[{Antonucci(1993)}]{antonucci_unified_1993}
Antonucci, R. 1993, ARA\&A, 31, 473

\bibitem[{Baes {et~al.}(2010)Baes, Clemens, Xilouris, Fritz, Cotton, Davies,
  Bendo, Bianchi, Cortese, de~Looze, Pohlen, Verstappen, B\"ohringer, Bomans,
  Boselli, Corbelli, Dariush, di~Serego~Alighieri, Fadda, {Garcia-Appadoo},
  Gavazzi, Giovanardi, Grossi, Hughes, Hunt, Jones, Madden, Pierini, Sabatini,
  Smith, Vlahakis, \& Zibetti}]{baes_herschel_2010}
Baes, M., Clemens, M., Xilouris, E.~M., {et~al.} 2010, A\&A, 518, L53

\bibitem[{Bassani {et~al.}(1999)Bassani, Dadina, Maiolino, Salvati, Risaliti,
  della Ceca, Matt, \& Zamorani}]{bassani_three-dimensional_1999}
Bassani, L., Dadina, M., Maiolino, R., {et~al.} 1999, ApJSS, 121, 473

\bibitem[{Bettoni {et~al.}(2003)Bettoni, Falomo, Fasano, \&
  Govoni}]{bettoni_black_2003}
Bettoni, D., Falomo, R., Fasano, G., \& Govoni, F. 2003, A\&A, 399, 869

\bibitem[{Bianchi {et~al.}(2009)Bianchi, Bonilla, Guainazzi, Matt, \&
  Ponti}]{bianchi_caixa:_2009}
Bianchi, S., Bonilla, N.~F., Guainazzi, M., Matt, G., \& Ponti, G. 2009, A\&A,
  501, 915

\bibitem[{Bianchi {et~al.}(2005)Bianchi, Guainazzi, Matt, Chiaberge, Iwasawa,
  Fiore, \& Maiolino}]{bianchi_search_2005}
Bianchi, S., Guainazzi, M., Matt, G., {et~al.} 2005, A\&A, 442, 185

\bibitem[{Blakeslee {et~al.}(2001)Blakeslee, Lucey, Barris, Hudson, \&
  Tonry}]{blakeslee_synthesis_2001}
Blakeslee, J.~P., Lucey, J.~R., Barris, B.~J., Hudson, M.~J., \& Tonry, J.~L.
  2001, MNRAS, 327, 1004

\bibitem[{Brandl {et~al.}(2006)Brandl, {Bernard-Salas}, Spoon, Devost, Sloan,
  Guilles, Wu, Houck, Weedman, Armus, Appleton, Soifer, Charmandaris, Hao,
  Higdon, \& Herter}]{brandl_mid-infrared_2006}
Brandl, B.~R., {Bernard-Salas}, J., Spoon, H. W.~W., {et~al.} 2006, ApJ, 653,
  1129

\bibitem[{Brightman \& Nandra(2011)}]{brightman_xmm-newton_2011}
Brightman, M. \& Nandra, K. 2011, MNRAS, 413, 1206

\bibitem[{Burtscher {et~al.}(2009)Burtscher, Jaffe, Raban, Meisenheimer,
  Tristram, \& R\"ottgering}]{burtscher_dust_2009-1}
Burtscher, L., Jaffe, W., Raban, D., {et~al.} 2009, The ApJL, 705, L53

\bibitem[{Cappellari {et~al.}(2009)Cappellari, Neumayer, Reunanen, van~der
  Werf, de~Zeeuw, \& Rix}]{cappellari_mass_2009}
Cappellari, M., Neumayer, N., Reunanen, J., {et~al.} 2009, MNRAS, 394, 660

\bibitem[{Cappellari {et~al.}(2002)Cappellari, Verolme, van~der Marel, Kleijn,
  Illingworth, Franx, Carollo, \& de~Zeeuw}]{cappellari_counterrotating_2002}
Cappellari, M., Verolme, E.~K., van~der Marel, R.~P., {et~al.} 2002, ApJ, 578,
  787

\bibitem[{Cappi {et~al.}(2006)Cappi, Panessa, Bassani, Dadina, Dicocco,
  Comastri, della Ceca, Filippenko, Gianotti, Ho, Malaguti, Mulchaey, Palumbo,
  Piconcelli, Sargent, Stephen, Trifoglio, \& Weaver}]{cappi_x-ray_2006}
Cappi, M., Panessa, F., Bassani, L., {et~al.} 2006, A\&A, 446, 459

\bibitem[{Carrillo {et~al.}(1999)Carrillo, Masegosa, {Dultzin-Hacyan}, \&
  Ordoñez}]{carrillo_multifrequency_1999}
Carrillo, R., Masegosa, J., {Dultzin-Hacyan}, D., \& Ordoñez, R. 1999, \rmxaa,
  35, 187

\bibitem[{Cohen {et~al.}(1999)Cohen, Walker, Carter, Hammersley, Kidger, \&
  Noguchi}]{cohen_spectral_1999}
Cohen, M., Walker, R.~G., Carter, B., {et~al.} 1999, AJ, 117, 1864

\bibitem[{Colina {et~al.}(2002)Colina, Gonzalez~Delgado, {Mas-Hesse}, \&
  Leitherer}]{colina_detection_2002}
Colina, L., Gonzalez~Delgado, R., {Mas-Hesse}, J.~M., \& Leitherer, C. 2002,
  ApJ, 579, 545

\bibitem[{{Diamond-Stanic} \& Rieke(2010)}]{diamond-stanic_effect_2010}
{Diamond-Stanic}, A.~M. \& Rieke, G.~H. 2010, ApJ, 724, 140

\bibitem[{{Diamond-Stanic} \& Rieke(2011)}]{diamond-stanic_relationship_2011}
{Diamond-Stanic}, A.~M. \& Rieke, G.~H. 2011, 1106.3565

\bibitem[{{Diamond-Stanic} {et~al.}(2009){Diamond-Stanic}, Rieke, \&
  Rigby}]{diamond-stanic_isotropic_2009}
{Diamond-Stanic}, A.~M., Rieke, G.~H., \& Rigby, J.~R. 2009, ApJ, 698, 623

\bibitem[{Dudik {et~al.}(2005)Dudik, Satyapal, Gliozzi, \&
  Sambruna}]{dudik_chandra_2005}
Dudik, R.~P., Satyapal, S., Gliozzi, M., \& Sambruna, R.~M. 2005, ApJ, 620, 113

\bibitem[{Dudik {et~al.}(2009)Dudik, Satyapal, \& Marcu}]{dudik_spitzer_2009}
Dudik, R.~P., Satyapal, S., \& Marcu, D. 2009, ApJ, 691, 1501

\bibitem[{Dullemond \& van Bemmel(2005)}]{dullemond_clumpy_2005}
Dullemond, C.~P. \& van Bemmel, I.~M. 2005, A\&A, 436, 47

\bibitem[{Elitzur \& Ho(2009)}]{elitzur_disappearance_2009}
Elitzur, M. \& Ho, L.~C. 2009, ApJ, 701, L91

\bibitem[{Elitzur \& Shlosman(2006)}]{elitzur_agn-obscuring_2006}
Elitzur, M. \& Shlosman, I. 2006, ApJ, 648, L101

\bibitem[{Falcke \& Markoff(2000)}]{falcke_jet_2000}
Falcke, H. \& Markoff, S. 2000, A\&A, 362, 113

\bibitem[{Ferrarese {et~al.}(1996)Ferrarese, Livio, Freedman, Saha, Stetson,
  Ford, Hill, \& Madore}]{ferrarese_discovery_1996}
Ferrarese, L., Livio, M., Freedman, W., {et~al.} 1996, ApJ, 468, L95

\bibitem[{Ferrarese \& Merritt(2000)}]{ferrarese_fundamental_2000}
Ferrarese, L. \& Merritt, D. 2000, ApJ, 539, L9

\bibitem[{Ferrarese {et~al.}(2007)Ferrarese, Mould, Stetson, Tonry, Blakeslee,
  \& Ajhar}]{ferrarese_discovery_2007}
Ferrarese, L., Mould, J.~R., Stetson, P.~B., {et~al.} 2007, ApJ, 654, 186

\bibitem[{Galliano {et~al.}(2008)Galliano, Alloin, Pantin, Granato, Delva,
  Silva, Lagage, \& Panuzzo}]{galliano_extremely_2008}
Galliano, E., Alloin, D., Pantin, E., {et~al.} 2008, A\&A, 492, 3

\bibitem[{Galliano {et~al.}(2005)Galliano, Alloin, Pantin, Lagage, \&
  Marco}]{galliano_mid-infrared_2005}
Galliano, E., Alloin, D., Pantin, E., Lagage, P.~O., \& Marco, O. 2005, A\&A,
  438, 803

\bibitem[{Gallimore {et~al.}(2010)Gallimore, Yzaguirre, Jakoboski, Stevenosky,
  Axon, Baum, Buchanan, Elitzur, Elvis, {O'Dea}, \&
  Robinson}]{gallimore_infrared_2010}
Gallimore, J.~F., Yzaguirre, A., Jakoboski, J., {et~al.} 2010, ApJSS, 187, 172

\bibitem[{Gandhi {et~al.}(2009)Gandhi, Horst, Smette, H\"onig, Comastri, Gilli,
  Vignali, \& Duschl}]{gandhi_resolving_2009}
Gandhi, P., Horst, H., Smette, A., {et~al.} 2009, A\&A, 502, 457

\bibitem[{{Garcia-Rissmann} {et~al.}(2005){Garcia-Rissmann}, Vega, Asari,
  Cid~Fernandes, Schmitt, Gonz\'alez~Delgado, \&
  {Storchi-Bergmann}}]{garcia-rissmann_atlas_2005}
{Garcia-Rissmann}, A., Vega, L.~R., Asari, N.~V., {et~al.} 2005, MNRAS, 359,
  765

\bibitem[{Gebhardt {et~al.}(2011)Gebhardt, Adams, Richstone, Lauer, Faber,
  Gultekin, Murphy, \& Tremaine}]{gebhardt_black-hole_2011}
Gebhardt, K., Adams, J., Richstone, D., {et~al.} 2011, 1101.1954

\bibitem[{Gebhardt {et~al.}(2000)Gebhardt, Bender, Bower, Dressler, Faber,
  Filippenko, Green, Grillmair, Ho, Kormendy, Lauer, Magorrian, Pinkney,
  Richstone, \& Tremaine}]{gebhardt_relationship_2000}
Gebhardt, K., Bender, R., Bower, G., {et~al.} 2000, ApJ, 539, L13

\bibitem[{{Gonz\'alez-Mart\'in}
  {et~al.}(2009{\natexlab{a}}){Gonz\'alez-Mart\'in}, Masegosa, M\'arquez, \&
  Guainazzi}]{gonzalez-martin_fitting_2009}
{Gonz\'alez-Mart\'in}, O., Masegosa, J., M\'arquez, I., \& Guainazzi, M.
  2009{\natexlab{a}}, ApJ, 704, 1570

\bibitem[{{Gonz\'alez-Mart\'in}
  {et~al.}(2009{\natexlab{b}}){Gonz\'alez-Mart\'in}, Masegosa, M\'arquez,
  Guainazzi, \& {Jim\'enez-Bail\'on}}]{gonzalez-martin_x-ray_2009}
{Gonz\'alez-Mart\'in}, O., Masegosa, J., M\'arquez, I., Guainazzi, M., \&
  {Jim\'enez-Bail\'on}, E. 2009{\natexlab{b}}, A\&A, 506, 1107

\bibitem[{Gorjian {et~al.}(2004)Gorjian, Werner, Jarrett, Cole, \&
  Ressler}]{gorjian_10_2004}
Gorjian, V., Werner, M.~W., Jarrett, T.~H., Cole, D.~M., \& Ressler, M.~E.
  2004, ApJ, 605, 156

\bibitem[{Goulding \& Alexander(2009)}]{goulding_towards_2009}
Goulding, A.~D. \& Alexander, D.~M. 2009, MNRAS, 398, 1165

\bibitem[{Grier {et~al.}(2011)Grier, Mathur, Ghosh, \&
  Ferrarese}]{grier_discovery_2011}
Grier, C.~J., Mathur, S., Ghosh, H., \& Ferrarese, L. 2011, ApJ, 731, 60

\bibitem[{Grossan(2004)}]{grossan_high_2004}
Grossan, B. 2004, astro-ph/0405190

\bibitem[{Groves {et~al.}(2006)Groves, Dopita, \&
  Sutherland}]{groves_infrared_2006}
Groves, B., Dopita, M., \& Sutherland, R. 2006, A\&A, 458, 405

\bibitem[{G\"ultekin {et~al.}(2009)G\"ultekin, Richstone, Gebhardt, Lauer,
  Tremaine, Aller, Bender, Dressler, Faber, Filippenko, Green, Ho, Kormendy,
  Magorrian, Pinkney, \& Siopis}]{gueltekin_m-_2009}
G\"ultekin, K., Richstone, D.~O., Gebhardt, K., {et~al.} 2009, ApJ, 698, 198

\bibitem[{Haas {et~al.}(2007)Haas, Siebenmorgen, Pantin, Horst, Smette,
  K\"aufl, Lagage, \& Chini}]{haas_visir_2007}
Haas, M., Siebenmorgen, R., Pantin, E., {et~al.} 2007, A\&A, 473, 369

\bibitem[{Haas {et~al.}(2005)Haas, Siebenmorgen, Schulz, Kr\"ugel, \&
  Chini}]{haas_spitzer_2005}
Haas, M., Siebenmorgen, R., Schulz, B., Kr\"ugel, E., \& Chini, R. 2005, A\&A,
  442, L39

\bibitem[{Hadfield \& Crowther(2006)}]{hadfield_how_2006}
Hadfield, L.~J. \& Crowther, P.~A. 2006, MNRAS, 368, 1822

\bibitem[{Hardcastle {et~al.}(2009)Hardcastle, Evans, \&
  Croston}]{hardcastle_active_2009}
Hardcastle, M.~J., Evans, D.~A., \& Croston, J.~H. 2009, MNRAS, 396, 1929

\bibitem[{{Hern\'an-Caballero} \&
  Hatziminaoglou(2011)}]{hernan-caballero_atlas_2011}
{Hern\'an-Caballero}, A. \& Hatziminaoglou, E. 2011, MNRAS, 414, 500

\bibitem[{Ho(2008)}]{ho_nuclear_2008}
Ho, L.~C. 2008, ARA\&A, 46, 475

\bibitem[{Ho {et~al.}(2001)Ho, Feigelson, Townsley, Sambruna, Garmire, Brandt,
  Filippenko, Griffiths, Ptak, \& Sargent}]{ho_detection_2001}
Ho, L.~C., Feigelson, E.~D., Townsley, L.~K., {et~al.} 2001, ApJ, 549, L51

\bibitem[{Ho {et~al.}(1995)Ho, Filippenko, \& Sargent}]{ho_search_1995}
Ho, L.~C., Filippenko, A.~V., \& Sargent, W.~L. 1995, ApJSS, 98, 477

\bibitem[{Ho {et~al.}(1997)Ho, Filippenko, \& Sargent}]{ho_search_1997-1}
Ho, L.~C., Filippenko, A.~V., \& Sargent, W. L.~W. 1997, ApJSS, 112, 315

\bibitem[{Ho {et~al.}(2009)Ho, Greene, Filippenko, \& Sargent}]{ho_search_2009}
Ho, L.~C., Greene, J.~E., Filippenko, A.~V., \& Sargent, W. L.~W. 2009, ApJSS,
  183, 1

\bibitem[{Ho \& Ulvestad(2001)}]{ho_radio_2001}
Ho, L.~C. \& Ulvestad, J.~S. 2001, ApJSS, 133, 77

\bibitem[{H\"onig \& Beckert(2007)}]{hoenig_active_2007}
H\"onig, S.~F. \& Beckert, T. 2007, MNRAS, 380, 1172

\bibitem[{H\"onig {et~al.}(2006)H\"onig, Beckert, Ohnaka, \&
  Weigelt}]{hoenig_radiative_2006}
H\"onig, S.~F., Beckert, T., Ohnaka, K., \& Weigelt, G. 2006, A\&A, 452, 459

\bibitem[{H\"onig \& Kishimoto(2010)}]{hoenig_dusty_2010}
H\"onig, S.~F. \& Kishimoto, M. 2010, A\&A, 523, 27

\bibitem[{H\"onig {et~al.}(2010)H\"onig, Kishimoto, Gandhi, Smette, Asmus,
  Duschl, Polletta, \& Weigelt}]{hoenig_dusty_2010-1}
H\"onig, S.~F., Kishimoto, M., Gandhi, P., {et~al.} 2010, A\&A, 515, 23

\bibitem[{Horst {et~al.}(2009)Horst, Duschl, Gandhi, \&
  Smette}]{horst_mid-infrared_2009}
Horst, H., Duschl, W.~J., Gandhi, P., \& Smette, A. 2009, A\&A, 495, 137

\bibitem[{Horst {et~al.}(2008)Horst, Gandhi, Smette, \&
  Duschl}]{horst_mid_2008}
Horst, H., Gandhi, P., Smette, A., \& Duschl, W.~J. 2008, A\&A, 479, 389

\bibitem[{Horst {et~al.}(2006)Horst, Smette, Gandhi, \&
  Duschl}]{horst_small_2006}
Horst, H., Smette, A., Gandhi, P., \& Duschl, W.~J. 2006, A\&A, 457, L17

\bibitem[{Hudaverdi {et~al.}(2006)Hudaverdi, Kunieda, Tanaka, Haba, Furuzawa,
  Tawara, \& Ercan}]{hudaverdi_overdensity_2006}
Hudaverdi, M., Kunieda, H., Tanaka, T., {et~al.} 2006, PASJ, 58, 931

\bibitem[{Humphrey \& Buote(2006)}]{humphrey_chandra_2006}
Humphrey, P.~J. \& Buote, D.~A. 2006, ApJ, 639, 136

\bibitem[{Isobe {et~al.}(1990)Isobe, Feigelson, Akritas, \&
  Babu}]{isobe_linear_1990}
Isobe, T., Feigelson, E.~D., Akritas, M.~G., \& Babu, G.~J. 1990, ApJ, 364, 104

\bibitem[{Jacoby {et~al.}(1992)Jacoby, Branch, Ciardullo, Davies, Harris,
  Pierce, Pritchet, Tonry, \& Welch}]{jacoby_critical_1992}
Jacoby, G.~H., Branch, D., Ciardullo, R., {et~al.} 1992, PASP, 104, 599

\bibitem[{Jaffe {et~al.}(2004)Jaffe, Meisenheimer, R\"ottgering, Leinert,
  Richichi, Chesneau, {Fraix-Burnet}, {Glazenborg-Kluttig}, Granato, Graser,
  Heijligers, K\"ohler, Malbet, Miley, Paresce, Pel, Perrin, Przygodda,
  Schoeller, Sol, Waters, Weigelt, Woillez, \& de~Zeeuw}]{jaffe_central_2004}
Jaffe, W., Meisenheimer, K., R\"ottgering, H. J.~A., {et~al.} 2004, Nature,
  429, 47

\bibitem[{Jensen {et~al.}(2003)Jensen, Tonry, Barris, Thompson, Liu, Rieke,
  Ajhar, \& Blakeslee}]{jensen_measuring_2003}
Jensen, J.~B., Tonry, J.~L., Barris, B.~J., {et~al.} 2003, ApJ, 583, 712

\bibitem[{{Jim\'enez-Bail\'on} {et~al.}(2003){Jim\'enez-Bail\'on},
  {Santos-Lle\'o}, {Mas-Hesse}, Guainazzi, Colina, Cerviño, \&
  Gonz\'alez~Delgado}]{jimenez-bailon_nuclear_2003}
{Jim\'enez-Bail\'on}, E., {Santos-Lle\'o}, M., {Mas-Hesse}, J.~M., {et~al.}
  2003, ApJ, 593, 127

\bibitem[{Kelly(2007)}]{kelly_aspects_2007}
Kelly, B.~C. 2007, ApJ, 665, 1489

\bibitem[{Kewley {et~al.}(2001)Kewley, Heisler, Dopita, \&
  Lumsden}]{kewley_optical_2001}
Kewley, L.~J., Heisler, C.~A., Dopita, M.~A., \& Lumsden, S. 2001, ApJSS, 132,
  37

\bibitem[{Kim {et~al.}(2006)Kim, Barkhouse, {Romero-Colmenero}, Green, Kim,
  Mossman, Schlegel, Silverman, Aldcroft, Anderson, Ivezic, Kashyap, Tananbaum,
  \& Wilkes}]{kim_chandra_2006}
Kim, D., Barkhouse, W.~A., {Romero-Colmenero}, E., {et~al.} 2006, ApJ, 644, 829

\bibitem[{Krabbe {et~al.}(2001)Krabbe, B\"oker, \&
  Maiolino}]{krabbe_n-band_2001}
Krabbe, A., B\"oker, T., \& Maiolino, R. 2001, ApJ, 557, 626

\bibitem[{Lagage {et~al.}(2004)Lagage, Pel, Authier, Belorgey, Claret, Doucet,
  Dubreuil, Durand, Elswijk, Girardot, K\"aufl, Kroes, Lortholary, Lussignol,
  Marchesi, Pantin, Peletier, Pirard, Pragt, Rio, Schoenmaker, Siebenmorgen,
  Silber, Smette, Sterzik, \& Veyssiere}]{lagage_successful_2004}
Lagage, P.~O., Pel, J.~W., Authier, M., {et~al.} 2004, The Messenger, 117, 12

\bibitem[{{LaMassa} {et~al.}(2010){LaMassa}, Heckman, Ptak, Martins, Wild, \&
  Sonnentrucker}]{lamassa_indicators_2010}
{LaMassa}, S.~M., Heckman, T.~M., Ptak, A., {et~al.} 2010, ApJ, 720, 786

\bibitem[{Leipski {et~al.}(2009)Leipski, Antonucci, Ogle, \&
  Whysong}]{leipski_spitzer_2009}
Leipski, C., Antonucci, R., Ogle, P., \& Whysong, D. 2009, ApJ, 701, 891

\bibitem[{Levenson {et~al.}(2006)Levenson, Heckman, Krolik, Weaver, \&
  \.{Z}ycki}]{levenson_penetrating_2006}
Levenson, N.~A., Heckman, T.~M., Krolik, J.~H., Weaver, K.~A., \& \.{Z}ycki,
  P.~T. 2006, ApJ, 648, 111

\bibitem[{Levenson {et~al.}(2009)Levenson, Radomski, Packham, Mason, Schaefer,
  \& Telesco}]{levenson_isotropic_2009}
Levenson, N.~A., Radomski, J.~T., Packham, C., {et~al.} 2009, ApJ, 703, 390

\bibitem[{Lewis \& Eracleous(2006)}]{lewis_black_2006}
Lewis, K.~T. \& Eracleous, M. 2006, ApJ, 642, 711

\bibitem[{Liu \& Bregman(2005)}]{liu_ultraluminous_2005}
Liu, J. \& Bregman, J.~N. 2005, ApJSS, 157, 59

\bibitem[{Lobban {et~al.}(2010)Lobban, Reeves, Porquet, Braito, Markowitz,
  Miller, \& Turner}]{lobban_evidence_2010}
Lobban, A.~P., Reeves, J.~N., Porquet, D., {et~al.} 2010, 1006.1318

\bibitem[{Lodato \& Bertin(2003)}]{lodato_non-keplerian_2003}
Lodato, G. \& Bertin, G. 2003, A\&A, 398, 517

\bibitem[{Lutz {et~al.}(2004)Lutz, Maiolino, Spoon, \&
  Moorwood}]{lutz_relation_2004}
Lutz, D., Maiolino, R., Spoon, H. W.~W., \& Moorwood, A. F.~M. 2004, A\&A, 418,
  465

\bibitem[{Maccarone {et~al.}(2003)Maccarone, Kundu, \&
  Zepf}]{maccarone_low-mass_2003}
Maccarone, T.~J., Kundu, A., \& Zepf, S.~E. 2003, ApJ, 586, 814

\bibitem[{Madore {et~al.}(1999)Madore, Freedman, Silbermann, Harding, Huchra,
  Mould, Graham, Ferrarese, Gibson, Han, Hoessel, Hughes, Illingworth, Phelps,
  Sakai, \& Stetson}]{madore_hubble_1999}
Madore, B.~F., Freedman, W.~L., Silbermann, N., {et~al.} 1999, ApJ, 515, 29

\bibitem[{Maoz(2007)}]{maoz_low-luminosity_2007}
Maoz, D. 2007, MNRAS, 377, 1696

\bibitem[{Markoff {et~al.}(2008)Markoff, Nowak, Young, Marshall, Canizares,
  Peck, Krips, Petitpas, Sch\"odel, Bower, Chandra, Ray, Muno, Gallagher,
  Hornstein, \& Cheung}]{markoff_results_2008}
Markoff, S., Nowak, M., Young, A., {et~al.} 2008, ApJ, 681, 905

\bibitem[{Mason {et~al.}(2007)Mason, Levenson, Packham, Elitzur, Radomski,
  Petric, \& Wright}]{mason_dust_2007}
Mason, R.~E., Levenson, N.~A., Packham, C., {et~al.} 2007, ApJ, 659, 241

\bibitem[{{McElroy}(1995)}]{mcelroy_catalog_1995}
{McElroy}, D.~B. 1995, ApJSS, 100, 105

\bibitem[{Mei {et~al.}(2007)Mei, Blakeslee, Côt\'e, Tonry, West, Ferrarese,
  Jord\'an, Peng, Anthony, \& Merritt}]{mei_acs_2007}
Mei, S., Blakeslee, J.~P., Côt\'e, P., {et~al.} 2007, ApJ, 655, 144

\bibitem[{Mel\'endez {et~al.}(2008)Mel\'endez, Kraemer, Armentrout, Deo,
  Crenshaw, Schmitt, Mushotzky, Tueller, Markwardt, \&
  Winter}]{melendez_new_2008}
Mel\'endez, M., Kraemer, S.~B., Armentrout, B.~K., {et~al.} 2008, ApJ, 682, 94

\bibitem[{Mor {et~al.}(2009)Mor, Netzer, \& Elitzur}]{mor_dusty_2009}
Mor, R., Netzer, H., \& Elitzur, M. 2009, ApJ, 705, 298

\bibitem[{Mullaney {et~al.}(2011)Mullaney, Alexander, Goulding, \&
  Hickox}]{mullaney_defining_2011}
Mullaney, J.~R., Alexander, D.~M., Goulding, A.~D., \& Hickox, R.~C. 2011,
  MNRAS, 474

\bibitem[{Murphy {et~al.}(2007)Murphy, Yaqoob, \&
  Terashima}]{murphy_monitoring_2007-1}
Murphy, K.~D., Yaqoob, T., \& Terashima, Y. 2007, ApJ, 666, 96

\bibitem[{Nagar {et~al.}(2005)Nagar, Falcke, \& Wilson}]{nagar_radio_2005}
Nagar, N.~M., Falcke, H., \& Wilson, A.~S. 2005, A\&A, 435, 521

\bibitem[{Narayan \& Yi(1994)}]{narayan_advection-dominated_1994}
Narayan, R. \& Yi, I. 1994, ApJ, 428, L13

\bibitem[{Nemmen {et~al.}(2006)Nemmen, {Storchi-Bergmann}, Yuan, Eracleous,
  Terashima, \& Wilson}]{nemmen_radiatively_2006}
Nemmen, R.~S., {Storchi-Bergmann}, T., Yuan, F., {et~al.} 2006, ApJ, 643, 652

\bibitem[{Nenkova {et~al.}(2002)Nenkova, Ivezi\'c, \&
  Elitzur}]{nenkova_dust_2002}
Nenkova, M., Ivezi\'c, v., \& Elitzur, M. 2002, ApJ, 570, L9

\bibitem[{Nenkova {et~al.}(2008)Nenkova, Sirocky, Nikutta, Ivezi\'c, \&
  Elitzur}]{nenkova_agn_2008-1}
Nenkova, M., Sirocky, M.~M., Nikutta, R., Ivezi\'c, v., \& Elitzur, M. 2008,
  ApJ, 685, 160

\bibitem[{Nicastro {et~al.}(2003)Nicastro, Martocchia, \&
  Matt}]{nicastro_lack_2003}
Nicastro, F., Martocchia, A., \& Matt, G. 2003, ApJ, 589, L13

\bibitem[{Panessa {et~al.}(2006)Panessa, Bassani, Cappi, Dadina, Barcons,
  Carrera, Ho, \& Iwasawa}]{panessa_x-ray_2006}
Panessa, F., Bassani, L., Cappi, M., {et~al.} 2006, A\&A, 455, 173

\bibitem[{Papadakis {et~al.}(2008)Papadakis, Ioannou, Brinkmann, \&
  Xilouris}]{papadakis_x-ray_2008}
Papadakis, I.~E., Ioannou, Z., Brinkmann, W., \& Xilouris, E.~M. 2008, A\&A,
  490, 995

\bibitem[{Pappa {et~al.}(2001)Pappa, Georgantopoulos, Stewart, \&
  Zezas}]{pappa_x-ray_2001}
Pappa, A., Georgantopoulos, I., Stewart, G.~C., \& Zezas, A.~L. 2001, MNRAS,
  326, 995

\bibitem[{Paturel {et~al.}(2003)Paturel, Petit, Prugniel, Theureau, Rousseau,
  Brouty, Dubois, \& Cambr\'esy}]{paturel_hyperleda._2003}
Paturel, G., Petit, C., Prugniel, P., {et~al.} 2003, A\&A, 412, 45

\bibitem[{Perlman {et~al.}(2007)Perlman, Mason, Packham, Levenson, Elitzur,
  Schaefer, Imanishi, Sparks, \& Radomski}]{perlman_mid-infrared_2007}
Perlman, E.~S., Mason, R.~E., Packham, C., {et~al.} 2007, ApJ, 663, 808

\bibitem[{Peterson {et~al.}(2004)Peterson, Ferrarese, Gilbert, Kaspi, Malkan,
  Maoz, Merritt, Netzer, Onken, Pogge, Vestergaard, \&
  Wandel}]{peterson_central_2004}
Peterson, B.~M., Ferrarese, L., Gilbert, K.~M., {et~al.} 2004, ApJ, 613, 682

\bibitem[{Press {et~al.}(1992)Press, Teukolsky, Vetterling, \&
  Flannery}]{press_numerical_1992}
Press, W.~H., Teukolsky, S.~A., Vetterling, W.~T., \& Flannery, B.~P. 1992,
  Numerical recipes in {FORTRAN.} The art of scientific computing

\bibitem[{Raban {et~al.}(2009)Raban, Jaffe, R\"ottgering, Meisenheimer, \&
  Tristram}]{raban_resolving_2009}
Raban, D., Jaffe, W., R\"ottgering, H., Meisenheimer, K., \& Tristram, K. R.~W.
  2009, MNRAS, 394, 1325

\bibitem[{{Ramos Almeida} {et~al.}(2009){Ramos Almeida}, Levenson,
  Rodr\'iguez~Espinosa, {Alonso-Herrero}, Asensio~Ramos, Radomski, Packham,
  Fisher, \& Telesco}]{ramos_almeida_infrared_2009}
{Ramos Almeida}, C., Levenson, N.~A., Rodr\'iguez~Espinosa, J.~M., {et~al.}
  2009, ApJ, 702, 1127

\bibitem[{Ranalli {et~al.}(2003)Ranalli, Comastri, \&
  Setti}]{ranalli_2-10_2003}
Ranalli, P., Comastri, A., \& Setti, G. 2003, A\&A, 399, 39

\bibitem[{Reunanen {et~al.}(2010)Reunanen, Prieto, \&
  Siebenmorgen}]{reunanen_vlt_2010}
Reunanen, J., Prieto, M.~A., \& Siebenmorgen, R. 2010, MNRAS, 402, 879

\bibitem[{Risaliti {et~al.}(2007)Risaliti, Elvis, Fabbiano, Baldi, Zezas, \&
  Salvati}]{risaliti_occultation_2007}
Risaliti, G., Elvis, M., Fabbiano, G., {et~al.} 2007, ApJ, 659, L111

\bibitem[{Sanders {et~al.}(2003)Sanders, Mazzarella, Kim, Surace, \&
  Soifer}]{sanders_iras_2003}
Sanders, D.~B., Mazzarella, J.~M., Kim, D., Surace, J.~A., \& Soifer, B.~T.
  2003, AJ, 126, 1607

\bibitem[{Sansom {et~al.}(2006)Sansom, {O'Sullivan}, Forbes, Proctor, \&
  Davis}]{sansom_x-ray_2006}
Sansom, A.~E., {O'Sullivan}, E., Forbes, D.~A., Proctor, R.~N., \& Davis, D.~S.
  2006, MNRAS, 370, 1541

\bibitem[{Satyapal {et~al.}(2005)Satyapal, Dudik, {O'Halloran}, \&
  Gliozzi}]{satyapal_link_2005}
Satyapal, S., Dudik, R.~P., {O'Halloran}, B., \& Gliozzi, M. 2005, ApJ, 633, 86

\bibitem[{Satyapal {et~al.}(2004)Satyapal, Sambruna, \&
  Dudik}]{satyapal_joint_2004}
Satyapal, S., Sambruna, R.~M., \& Dudik, R.~P. 2004, A\&A, 414, 825

\bibitem[{Schartmann {et~al.}(2008)Schartmann, Meisenheimer, Camenzind, Wolf,
  Tristram, \& Henning}]{schartmann_three-dimensional_2008}
Schartmann, M., Meisenheimer, K., Camenzind, M., {et~al.} 2008, A\&A, 482, 67

\bibitem[{Schwope {et~al.}(2000)Schwope, Hasinger, Lehmann, Schwarz, Brunner,
  Neizvestny, Ugryumov, Balega, Tr\"umper, \& Voges}]{schwope_rosat_2000}
Schwope, A., Hasinger, G., Lehmann, I., {et~al.} 2000, Astron. Nachr., 321, 1

\bibitem[{Shakura \& Sunyaev(1973)}]{shakura_black_1973}
Shakura, N.~I. \& Sunyaev, R.~A. 1973, A\&A, 24, 337

\bibitem[{Shi {et~al.}(2010)Shi, Rieke, Smith, Rigby, Hines, Donley, Schmidt,
  \& {Diamond-Stanic}}]{shi_unobscured_2010}
Shi, Y., Rieke, G.~H., Smith, P., {et~al.} 2010, ApJ, 714, 115

\bibitem[{Shu {et~al.}(2010{\natexlab{a}})Shu, Liu, \&
  Wang}]{shu_xmm-newton_2010}
Shu, X.~W., Liu, T., \& Wang, J.~X. 2010{\natexlab{a}}, ApJ, 722, 96

\bibitem[{Shu {et~al.}(2010{\natexlab{b}})Shu, Yaqoob, \&
  Wang}]{shu_cores_2010}
Shu, X.~W., Yaqoob, T., \& Wang, J.~X. 2010{\natexlab{b}}, ApJSS, 187, 581

\bibitem[{Siebenmorgen {et~al.}(2008)Siebenmorgen, Haas, Pantin, Kr\"ugel,
  Leipski, K\"aufl, Lagage, Moorwood, Smette, \&
  Sterzik}]{siebenmorgen_nuclear_2008}
Siebenmorgen, R., Haas, M., Pantin, E., {et~al.} 2008, A\&A, 488, 83

\bibitem[{Slee {et~al.}(1994)Slee, Sadler, Reynolds, \&
  Ekers}]{slee_parsecscale_1994}
Slee, O.~B., Sadler, E.~M., Reynolds, J.~E., \& Ekers, R.~D. 1994, MNRAS, 269,
  928

\bibitem[{Snijders {et~al.}(2006)Snijders, van~der Werf, Brandl, Mengel,
  Schaerer, \& Wang}]{snijders_subarcsecond_2006}
Snijders, L., van~der Werf, P.~P., Brandl, B.~R., {et~al.} 2006, ApJ, 648, L25

\bibitem[{Springob {et~al.}(2007)Springob, Masters, Haynes, Giovanelli, \&
  Marinoni}]{springob_sfi++._2007}
Springob, C.~M., Masters, K.~L., Haynes, M.~P., Giovanelli, R., \& Marinoni, C.
  2007, ApJSS, 172, 599

\bibitem[{Starling {et~al.}(2005)Starling, Page, {Branduardi-Raymont},
  Breeveld, Soria, \& Wu}]{starling_x-ray_2005}
Starling, R. L.~C., Page, M.~J., {Branduardi-Raymont}, G., {et~al.} 2005,
  MNRAS, 356, 727

\bibitem[{{Storchi-Bergmann} {et~al.}(2005){Storchi-Bergmann}, Nemmen,
  Spinelli, Eracleous, Wilson, Filippenko, \&
  Livio}]{storchi-bergmann_evidence_2005}
{Storchi-Bergmann}, T., Nemmen, R.~S., Spinelli, P.~F., {et~al.} 2005, ApJ,
  624, L13

\bibitem[{Terashima {et~al.}(2000)Terashima, Ho, \& Ptak}]{terashima_hard_2000}
Terashima, Y., Ho, L.~C., \& Ptak, A.~F. 2000, ApJ, 539, 161

\bibitem[{Tremaine {et~al.}(2002)Tremaine, Gebhardt, Bender, Bower, Dressler,
  Faber, Filippenko, Green, Grillmair, Ho, Kormendy, Lauer, Magorrian, Pinkney,
  \& Richstone}]{tremaine_slope_2002}
Tremaine, S., Gebhardt, K., Bender, R., {et~al.} 2002, ApJ, 574, 740

\bibitem[{Tristram {et~al.}(2009)Tristram, Raban, Meisenheimer, Jaffe,
  R\"ottgering, Burtscher, Cotton, Graser, Henning, Leinert, Lopez, Morel,
  Perrin, \& Wittkowski}]{tristram_parsec-scale_2009}
Tristram, K. R.~W., Raban, D., Meisenheimer, K., {et~al.} 2009, A\&A, 502, 67

\bibitem[{Tristram \& Schartmann(2011)}]{tristram_size-luminosity_2011}
Tristram, K. R.~W. \& Schartmann, M. 2011, A\&A, 531, 99

\bibitem[{Tully(1988)}]{tully_nearby_1988}
Tully, R.~B. 1988, Nearby galaxies catalog

\bibitem[{Tzanavaris \& Georgantopoulos(2007)}]{tzanavaris_searching_2007}
Tzanavaris, P. \& Georgantopoulos, I. 2007, A\&A, 468, 129

\bibitem[{van~der Wolk {et~al.}(2010)van~der Wolk, Barthel, Peletier, \&
  Pel}]{van_der_wolk_dust_2010}
van~der Wolk, G., Barthel, P.~D., Peletier, R.~F., \& Pel, J.~W. 2010, A\&A,
  511, 64

\bibitem[{Vasudevan \& Fabian(2007)}]{vasudevan_piecing_2007}
Vasudevan, R.~V. \& Fabian, A.~C. 2007, MNRAS, 381, 1235

\bibitem[{Vasudevan {et~al.}(2010)Vasudevan, Fabian, Gandhi, Winter, \&
  Mushotzky}]{vasudevan_power_2010}
Vasudevan, R.~V., Fabian, A.~C., Gandhi, P., Winter, L.~M., \& Mushotzky, R.~F.
  2010, MNRAS, 402, 1081

\bibitem[{Vasudevan {et~al.}(2009)Vasudevan, Mushotzky, Winter, \&
  Fabian}]{vasudevan_optical--x-ray_2009}
Vasudevan, R.~V., Mushotzky, R.~F., Winter, L.~M., \& Fabian, A.~C. 2009,
  MNRAS, 399, 1553

\bibitem[{{V\'eron-Cetty} \& V\'eron(2010)}]{veron-cetty_catalogue_2010}
{V\'eron-Cetty}, M. \& V\'eron, P. 2010, A\&A, 518, 10

\bibitem[{Wang \& Zhang(2007)}]{wang_unified_2007}
Wang, J. \& Zhang, E. 2007, ApJ, 660, 1072

\bibitem[{Willick {et~al.}(1997)Willick, Courteau, Faber, Burstein, Dekel, \&
  Strauss}]{willick_homogeneous_1997}
Willick, J.~A., Courteau, S., Faber, S.~M., {et~al.} 1997, ApJSS, 109, 333

\bibitem[{Winter {et~al.}(2010)Winter, Lewis, Koss, Veilleux, Keeney, \&
  Mushotzky}]{winter_optical_2010}
Winter, L.~M., Lewis, K.~T., Koss, M., {et~al.} 2010, ApJ, 710, 503

\bibitem[{Woo {et~al.}(2010)Woo, Treu, Barth, Wright, Walsh, Bentz, Martini,
  Bennert, Canalizo, Filippenko, Gates, Greene, Li, Malkan, Stern, \&
  Minezaki}]{woo_lick_2010}
Woo, J., Treu, T., Barth, A.~J., {et~al.} 2010, ApJ, 716, 269

\bibitem[{Woo \& Urry(2002)}]{woo_active_2002}
Woo, J. \& Urry, C.~M. 2002, ApJ, 579, 530

\bibitem[{Yuan(2007)}]{yuan_advection-dominated_2007}
Yuan, F. 2007, in , 95

\bibitem[{Yuan \& Cui(2005)}]{yuan_radio-x-ray_2005}
Yuan, F. \& Cui, W. 2005, ApJ, 629, 408

\bibitem[{Yuan {et~al.}(2009)Yuan, Yu, \& Ho}]{yuan_revisiting_2009}
Yuan, F., Yu, Z., \& Ho, L.~C. 2009, ApJ, 703, 1034

\bibitem[{Zhang {et~al.}(2009)Zhang, Soria, Zhang, Swartz, \&
  Liu}]{zhang_census_2009}
Zhang, W.~M., Soria, R., Zhang, S.~N., Swartz, D.~A., \& Liu, J.~F. 2009, ApJ,
  699, 281

\end{thebibliography}

\appendix

\section{Brighter AGN properties}
For completeness the properties of the AGN sample from G+09 relevant here are stated in Table~\ref{tab:AGN}.
\begin{table*}
\begin{minipage}[t]{\textwidth}
\caption{Luminosities and other properties for the G+09 minus LLAGN.
}

\centering 
\label{tab:AGN}      
\renewcommand{\footnoterule}{}  
\begin{tabular}{l c c c c c c c c c c}        
\hline\hline    
Object & Seyfert type & $\log L_{2-10\,\mathrm{keV}}$ & $\log \lambda L_{12.3\,\mu\mathrm{m}}$ & $\log M_\mathrm{BH}$ & Ref.& $\log \lambda$\\
       &              & [erg/s]                      & [erg/s]                                & [$M_\odot$]          &     &        \\
\hline
Fairall9	&	1.2	&	43.87	$\pm$	0.15	&	44.51	$\pm$	0.03	&	7.91	&	8	&	-1.0	\\
NGC526A	&	1.9	&	43.14	$\pm$	0.10	&	43.70	$\pm$	0.08	&	8.10	&	2	&	-1.9	\\
SWIFTJ0138.6-4001	&	2	&	43.59	$\pm$	0.30	&	43.10	$\pm$	0.04	&	9.68	&	9	&	-3.6	\\
Mrk590	&	1	&	43.61	$\pm$	0.25	&	43.55	$\pm$	0.06	&	7.66	&	10	&	-1.3	\\
NGC1068	&	1h	&	43.40	$\pm$	0.30	&	43.80	$\pm$	0.03	&	6.93	&	4	&	-0.6	\\
ESO209-G012	&	1.5	&	43.65	$\pm$	0.15	&	44.22	$\pm$	0.06	&	\dots	&	\dots	&	\dots	\\
NGC3081	&	1h	&	42.60	$\pm$	0.30	&	42.64	$\pm$	0.04	&	7.31	&	2	&	-1.9	\\
ESO263-G013	&	2	&	43.30	$\pm$	0.40	&	43.42	$\pm$	0.02	&	\dots	&	\dots	&	\dots	\\
NGC3281	&	2	&	43.30	$\pm$	0.15	&	43.76	$\pm$	0.02	&	7.72	&	2	&	-1.4	\\
NGC3393	&	2	&	43.04	$\pm$	0.30	&	42.90	$\pm$	0.02	&	8.09	&	2	&	-2.4	\\
NGC3783	&	1.5	&	43.21	$\pm$	0.15	&	43.52	$\pm$	0.05	&	7.45	&	10	&	-1.3	\\
NGC4388	&	1h	&	42.24	$\pm$	0.30	&	42.45	$\pm$	0.03	&	7.10	&	3	&	-2.0	\\
NGC4507	&	1h	&	43.30	$\pm$	0.15	&	43.67	$\pm$	0.04	&	8.39	&	9	&	-2.2	\\
NGC4593	&	1	&	42.93	$\pm$	0.20	&	43.18	$\pm$	0.08	&	6.97	&	10	&	-1.2	\\
IC3639	&	1h	&	43.62	$\pm$	0.50	&	43.51	$\pm$	0.02	&	6.75	&	2	&	-0.4	\\
ESO323-G032	&	1.9	&	42.43	$\pm$	0.40	&	42.96	$\pm$	0.04	&	\dots	&	2	&	\dots	\\
NGC4992	&	2	&	43.23	$\pm$	0.30	&	43.36	$\pm$	0.02	&	8.56	&	9	&	-2.5	\\
IRAS13197-1627	&	1h	&	42.78	$\pm$	0.20	&	44.07	$\pm$	0.03	&	8.28	&	9	&	-2.1	\\
NGC5135	&	2	&	43.00	$\pm$	0.50	&	43.06	$\pm$	0.05	&	7.30	&	2	&	-1.5	\\
MGC-06-30-15	&	1.5	&	42.74	$\pm$	0.20	&	43.07	$\pm$	0.06	&	6.73	&	2	&	-1.1	\\
NGC5728	&	1.9	&	42.80	$\pm$	0.30	&	42.69	$\pm$	0.03	&	8.20	&	5	&	-2.7	\\
NGC5995	&	2	&	43.54	$\pm$	0.15	&	44.11	$\pm$	0.06	&	7.11	&	7	&	-0.5	\\
ESO138-G001	&	2	&	43.00	$\pm$	0.30	&	43.54	$\pm$	0.02	&	\dots	&	\dots	&	\dots	\\
NGC6300	&	2	&	42.30	$\pm$	0.30	&	42.69	$\pm$	0.02	&	6.77	&	2	&	-1.5	\\
ESO103-G035	&	2	&	43.44	$\pm$	0.15	&	43.74	$\pm$	0.03	&	7.73	&	2	&	-1.4	\\
ESO141-G055	&	1	&	43.90	$\pm$	0.15	&	44.04	$\pm$	0.11	&	7.91	&	7	&	-1.2	\\
NGC6814	&	1.5	&	42.12	$\pm$	0.40	&	42.07	$\pm$	0.02	&	7.28	&	8	&	-2.4	\\
Mrk509	&	1.5	&	44.10	$\pm$	0.15	&	44.18	$\pm$	0.07	&	7.86	&	8	&	-1.0	\\
PKS2048-57	&	1h	&	42.84	$\pm$	0.20	&	43.82	$\pm$	0.05	&	7.74	&	8	&	-1.7	\\
PG2130+099	&	1.5	&	43.65	$\pm$	0.20	&	44.53	$\pm$	0.07	&	7.91	&	6	&	-1.1	\\
NGC7172	&	2	&	42.76	$\pm$	0.40	&	42.79	$\pm$	0.07	&	7.64	&	2	&	-2.1	\\
3C445	&	1.5	&	44.26	$\pm$	0.15	&	44.50	$\pm$	0.06	&	8.15	&	2	&	-1.0	\\
NGC7314	&	1h	&	42.20	$\pm$	0.15	&	41.96	$\pm$	0.12	&	7.84	&	9	&	-3.0	\\
NGC7469	&	1.5	&	43.15	$\pm$	0.10	&	43.92	$\pm$	0.03	&	7.06	&	10	&	-0.8	\\
NGC7674	&	1h	&	44.56	$\pm$	0.50	&	44.31	$\pm$	0.03	&	7.59	&	3	&	-0.4	\\
NGC7679	&	2	&	42.52	$\pm$	0.15	&	42.82	$\pm$	0.15	&	6.35	&	3	&	-0.9	\\

\hline                                   
\end{tabular}
\end{minipage}
\newline 
{\it -- Notes:} 
References for the black hole masses. 
(1) \cite{bettoni_black_2003}; 
(2) \cite{garcia-rissmann_atlas_2005}; 
(3) Hyperleda database \citep{paturel_hyperleda._2003}; 
(4) \cite{lodato_non-keplerian_2003}; 
(5) \cite{lewis_black_2006}; 
(6) \cite{mcelroy_catalog_1995};
(7) \cite{peterson_central_2004};
(8) \cite{wang_unified_2007};
(9) \cite{woo_active_2002}; 
(10) \cite{winter_optical_2010}; 
(11) \cite{woo_lick_2010}. 
$\log \lambda = \log (L_\mathrm{Bol} / L_\mathrm{Edd})$ is the Eddington ratio.

\end{table*}

\section{Additional notes on individual objects}\label{app:prop}

\subsection{\object{IC 1459}}
Detailed X-ray modeling of {\it Chandra} observations of this object are presented in \cite{gonzalez-martin_x-ray_2009}. \cite{satyapal_joint_2004} give comparable model parameters for the same data set. For the X-ray luminosity the mean of both works is used.  There is no [OIII] $\lambda 5007\,$\AA\, or [OIV] $\lambda 25.89\,\mu$m luminosity measurement in the literature. 
The X-ray data suggest that the object is not Compton-thick. In addition, a compact radio core was detected by \cite{slee_parsecscale_1994}, confirming the AGN nature of this object. Thus, the non-detection with VISIR may be explainable by the bad weather conditions during the observations of IC 1459.

\subsection{\object{NGC 676}}
The only redshift-independent distance measurement for NGC 676 available in the literature is by \cite{tully_nearby_1988} employing the Tully-Fisher relation, and thus is used in this work. This object is included in the Palomar optical spectroscopic survey \citep{ho_search_1995}. But it was not detected in radio by \cite{ho_radio_2001}. The used X-ray data based on {\it XMM-Newton} is published in \cite{panessa_x-ray_2006}. They interpret this object as Compton-thick in accordance to \cite{akylas_xmm-newton_2009} based on the same data. This is as well supported by the relatively strong [OIII] emission \citep{panessa_x-ray_2006}. Unfortunately, there is no {\it Spitzer} data available, and thus no [OIV] measurement.

\subsection{\object{NGC 1052}}
The X-ray properties of NGC 1052 are described in \cite{gonzalez-martin_x-ray_2009} based on {\it Chandra} data. The derived X-ray luminosity is very similar to the one given in \cite{satyapal_joint_2004} for the same observation after correction for the different distances. The mean of both is used in this work. 
Counting as a prototypical LINER, NGC 1052 is surprisingly bright and obscured in X-rays. While the [OIII] luminosity is in agreement with the X-ray luminosity \citep{gonzalez-martin_fitting_2009}, there has been no [OIV] emission line detected \citep{satyapal_joint_2004}.
Very recently \cite{brightman_xmm-newton_2011} published high S/N {\it XMM-Newton} data of this object stating a higher intrinsic $\log L_\mathrm{2-10keV} = 41.5$, which would bring this object closer to the correlation. 

\subsection{\object{NGC 1097}}
The X-ray luminosity is taken from \cite{nemmen_radiatively_2006} using {\it Chandra} observations.
It is interesting to note that the measured [OIV] brightness would imply a much higher X-ray luminosity ($\log L_\mathrm{X} \sim 42.2$) according to the  correlation found between both \citep{diamond-stanic_isotropic_2009} while the [OIII] luminosity implies a lower value ($\log L_\mathrm{X} \sim 40.0$) using the [OIII]:X-ray correlation \citep{panessa_x-ray_2006}.
At larger scale ($\sim 10\,\arcsec$) the well-known starburst ring is evident in both filters but can only be partly seen in the small VISIR field of view.
The infrared properties of the nucleus have already been investigated in detail by \cite{mason_dust_2007}. 
They find the absence of PAH 3.3\,$\mu$m emission at subarcsec scale. which seems to be typical for strong radiation fields. 
On the other hand, attempts to fit the infrared data with torus models failed, which led them to the conclusion that the torus is absent or weak.
Instead, a nuclear star-forming region, co-existing with the LLAGN, is assumed to dominate the observed MIR emission. 
The likely PAH 11.3\,$\mu$m emission in the VISIR data indicates that the nucleus is indeed dominated by emission coming from such a star-forming region. 
This also matches with the finding of \cite{storchi-bergmann_evidence_2005} of UV evidence for a nuclear star cluster inside the innermost 9 parsec of NGC 1097. 
But on the other hand, the MIR-excess relative to the luminosity correlation is consistent with the co-existence of star-forming region and a torus inside the nucleus of NGC 1097. 
For further disentangling of both components additional MIR observations are needed.

\subsection{\object{NGC 1386}}
NGC 1386 is one of the objects for which the different methods used to determine the distance show the largest discrepancies ($\sim 50\,\%$ variation). As this object is very likely Compton-thick it is difficult to derive a reliable X-ray luminosity. Here the approach of \cite{levenson_penetrating_2006} is followed and the luminosity is estimated by using the [OIII] line strength. The derived value is consistent with the Fe K$\alpha$ emission line properties observed by Chandra and analyzed in the same work. It also agrees with the [OIV] brightness stated in \cite{diamond-stanic_isotropic_2009} and \cite{lamassa_indicators_2010}. Thus, all available evidence favors a Compton-thick nature for this object.

\subsection{\object{NGC 1404}}
This elliptical galaxy is falling into the center of the Fornax cluster leading to large differences in the different distance measurement methods. It is classified as NELG in the optical. No evidence for AGN activity was found in the literature. Nevertheless, it is relatively bright in X-rays with a luminosity of $\log L_\mathrm{X} \sim 41$ reported by \cite{kim_chandra_2006} based on {\it Chandra} observations. As this value is calculated for the whole galaxy including also soft X-ray emission it has to be regarded as an upper limit for any Compton-thin AGN in this object. Very recently, \cite{grier_discovery_2011} re-examined the {\it Chandra} data and find a clear detection of a nuclear point source  with $\log L_\mathrm{X} \sim 40$ making a LLAGN in this object likely. 
However, in this work NGC 1404 is treated as non-AGN galaxy and comparison object only.

\subsection{\object{NGC 1566}}
Similar to NGC 1386, NGC 1566 is showing huge discrepancies between the different distance measurement methods. The only published X-ray data of this objects is from {\it ROSAT} \citep{liu_ultraluminous_2005}, where the emphasis was on off-nuclear ultra-luminous X-ray sources. Thus, the nuclear point-source was fitted only by a simple power-law. In the lack of any other available data, these values are used nevertheless. This X-ray luminosity is in good agreement with the predicted values from the [OIII] and [OIV] emission line measurements stated in \cite{diamond-stanic_isotropic_2009}.

\subsection{\object{NGC 1667}}
\cite{bianchi_search_2005} finds an X-ray flux variation of a factor of 100 for this object and interprets the data such, that the object is Compton-thick. Unfortunately, they give no intrinsic X-ray luminosity or flux derived from the used {\it XMM-Newton} data. Thus, here the value of \cite{panessa_x-ray_2006} based on {\it ASCA} observations is used. The latter authors as well interpret the data in favor of Compton-thickness. This is supported by the appearance of a strong Fe K$\alpha$ line \citep{bassani_three-dimensional_1999}.  On the other hand, the same {\it ASCA} data were originally used by \cite{pappa_x-ray_2001} for a Compton-thin fit, giving a two orders of magnitude lower X-ray luminosity for a similar observed flux. The stated [OIII] measurements in \cite{panessa_x-ray_2006} support a Compton-thick nature. The [OIV] emission \citep{diamond-stanic_isotropic_2009,lamassa_indicators_2010} predicts as well an intrinsic X-ray luminosity similar to the one for the Compton-thick case. The VISIR observations support this scenario with NGC 1667 being much closer to the MIR-X-ray correlation than in the unabsorbed scenario.

\subsection{\object{NGC 3312}}
NGC 3312 was classified as LINER by \cite{carrillo_multifrequency_1999}. The X-ray data are taken from \cite{hudaverdi_overdensity_2006} which describes {\it XMM-Newton} observations of this object. No other X-ray data are available as well as no {\it Spitzer} or [OIII] observations.

\subsection{\object{NGC 4235}}
{\it XMM-Newton} data analyzed in \cite{papadakis_x-ray_2008} and \cite{bianchi_caixa:_2009} yielding similar results is used here. \cite{panessa_x-ray_2006} derived a 0.6\,dex higher luminosity but based on {\it ASCA} observations, which are treated with a lower priority due to less spatial resolution. It is noteworthy that the properties of this type 1 LLAGN are in good agreement with those of brighter objects, and thus it is likely that NGC 4235 accretes in radiatively efficient mode as well \citep{papadakis_x-ray_2008}.

\subsection{\object{NGC 4261} - 3C 270}
\cite{gonzalez-martin_x-ray_2009} and \cite{satyapal_link_2005} arrive at very similar X-ray luminosities based on the same {\it Chandra} observation. 
It is interesting to note, that \cite{van_der_wolk_dust_2010} reports a non-detection with VISIR in the SiC filter for this source. 
These observations were re-analyzed by us and a weak but clear detection was found. 
The measured flux matches the new observations presented here, and is 6 times higher than their reported upper limit (2\,mJy). 
Additional evidence for MIR emission of hot dust in the nucleus of this galaxy has been found by \cite{leipski_spitzer_2009}. 
Thus, a dusty torus is likely present in NGC 4261, which is in agreement with the fact that the object aligns well with the MIR-X-ray correlation.

\subsection{\object{NGC 4472}}
This object was included in \cite{horst_mid_2008} erroneously because the nuclear X-ray luminosity was ambiguous due to many off-nuclear X-ray sources. Only an upper limit for the AGN ($\log L_\mathrm{X} < 39.32$) derived from \textit{Chandra} observations is given in \cite{panessa_x-ray_2006}. On the other hand, in a very detailed analysis \cite{maccarone_low-mass_2003} found a nuclear X-ray source with $\log L_\mathrm{X} \sim 39$ in the same data. This is in very good agreement with the [OIII] emission line brightness stated in \cite{diamond-stanic_isotropic_2009} while the [OIV] data indicate a much higher luminosity ($\log L_\mathrm{X} \sim 41.4$). But the latter seems to be the only indication of Compton-thickness which is also not supported by the non-detection with VISIR. NGC 4472 might be a so-called ``true'' Seyfert 2 candidate -- meaning that the broad emission lines are intrinsically absent rather than obscured by a torus \citep{nicastro_lack_2003}.
This was discussed in \cite{akylas_xmm-newton_2009}, but the \textit{XMM-Newton} data presented there is heavily contaminated by off-nuclear emission and thus unreliable.  Unfortunately, no conclusion about the existence of a torus in NGC 4472 can be drawn in this work

\subsection{\object{NGC 4486} - M 87}
Here the mean of the X-ray luminosities of \cite{gonzalez-martin_x-ray_2009}  and \cite{satyapal_joint_2004} is used. Both were using the same {\it Chandra} data and state very comparable results. The flux seen in the {\it Spitzer} IRS spectrum matches well with the VISIR data. Interestingly, \cite{perlman_mid-infrared_2007} explains all observed MIR emission with synchrotron emission by the jet and conclude that there is probably no torus present in NGC 4486. Very recent {\it Herschel}/PACS and SPIRE observations  \citep{baes_herschel_2010} show the absence of significant thermal components in the MIR to far-infrared SED on large scale ($> 10\arcsec$): a single power law can roughly fit the overall SED from MIR all way to the radio with a slope of $\sim-0.7$. But due to the comparably low spatial resolution no conclusion about the presence of a torus can be drawn from this data.

\subsection{\object{NGC 4594} - M 104}
 X-ray properties are taken from \cite{gonzalez-martin_x-ray_2009} and \cite{ho_detection_2001}  based on {\it Chandra} with marginally different results. This object was detected in the N-band by \cite{grossan_high_2004} with Keck LWS. The reported nuclear flux at 10.2\,$\mu$m (11.1\,mJy) is consistent with the upper limit derived here and would place this object close to the correlation. The \textit{Spitzer}/IRS spectrum is very noisy and the much higher flux implies heavy non-nuclear contamination at this lower spatial resolution. 

Similar to NGC 4472, NGC 4594 is also is a so-called ``true'' Sy 2 candidate and was investigated in detail in
\cite{shi_unobscured_2010}. They found strong silicate emission indicating that this object possesses a face-on dusty torus. 
The combined multiwavelength properties, including X-ray and optical/UV imply that NGC 4594 indeed lacks broad emission lines in contradiction of the simple unification models.

\subsection{\object{NGC 4698}}
\cite{cappi_x-ray_2006} and \cite{gonzalez-martin_x-ray_2009} give X-ray spectral fits with low absorption differing by 0.4\,dex derived from \textit{XMM-Newton} and \textit{Chandra} data ($\log L_\mathrm{X} = 39.39$ and 38.97), respectively. On the other hand, \cite{gonzalez-martin_fitting_2009} interpret NGC 4698 as Compton-thick using the same \textit{Chandra} data. This conclusion is based on the [OIII] to X-ray flux ratio. Similarly high [OIII] fluxes are stated in \cite{panessa_x-ray_2006} and \cite{diamond-stanic_isotropic_2009}. In addition, a high X-ray obscuration is consistent with the [OIV] emission line brightness of the latter work. Thus, the object is very likely Compton-thick but with a large luminosity uncertainty. 
However,  MIR upper limit of NGC 4698 is consistent with both the Compton-thin or thick scenario.

\subsection{\object{NGC 5363}}
Initial X-ray studies, \cite{gonzalez-martin_x-ray_2009} and \cite{sansom_x-ray_2006}, applied unobscured spectral fits to the available {\it XMM-Newton} observations and calculate a low X-ray luminosity ($\log L_\mathrm{X} = 39.68$ and 40.30) for NGC 5363. 
However, in the most recent analysis \citep{gonzalez-martin_fitting_2009} give a much higher luminosity from the same data but with a Compton-thick model. This is supported by the [OIII] brightness and a relatively flat spectrum above 2\,keV. Unfortunately, there is no {\it Spitzer}/IRS spectrum and thus no [OIV] measurement available. Here the Compton-thick fit will be adopted. 

\subsection{\object{NGC 5813}}
Here the X-ray luminosity of \cite{gonzalez-martin_x-ray_2009} based on {\it Chandra} is used. The X-ray image shows extremely diffuse soft emission. But no hard emission core is evident, which contradicts the existence of an AGN in this object. On the other hand, it could be an Compton-thick object as argued in \cite{gonzalez-martin_fitting_2009}. But the [OIII] luminosity given in the same work predicts a only slightly higher X-ray luminosity than in the Compton-thin fit. The AGN nature is supported by the detection of a compact radio core by \cite{nagar_radio_2005}. But even by adopting the Compton-thick luminosity ($\log L_\mathrm{X} = 40.55$), the non-detection with VISIR is not surprising and the derived upper limit does not constrain the correlation.

\subsection{\object{NGC 7213}}
X-ray measurements by {\it XMM-Newton} \citep[e.g.][]{starling_x-ray_2005}, {\it Chandra} \citep{shu_cores_2010} and {\it Suzaku} \citep{lobban_evidence_2010} available in the literature give very similar X-ray luminosities. Here, the mean value is used. As these come from different satellites and epochs, it seems unlikely that this object varies strongly in the X-rays. Thus, a smaller uncertainty is adopted for this object (0.1\,dex).

\subsection{\object{NGC 7590}}
{\it XMM-Newton} data reveals that the X-ray emission of NGC 7590 is dominated by an off-nuclear source and extended soft emission from the host galaxy \citep{shu_xmm-newton_2010}. Both dominates the older {\it ASCA} observation by \cite{bassani_three-dimensional_1999}  with less spatial resolution. Unfortunately, for this object there is no  {\it Chandra} observation available yet. Here the estimated nuclear hard X-ray luminosity of \cite{shu_xmm-newton_2010} is used. But note that based on the [OIII] to 2-10\,keV flux ratio the object is likely Compton-thick, which would imply a much higher intrinsic value. This is in contradiction to the hypothesis, that this object is a ``true'' Sy 2 candidate \citep[compare also][]{shi_unobscured_2010}. The upper MIR flux limit derived from VISIR observation does not help to constrain its nature as even a 2 orders of magnitude higher intrinsic 2-10\,keV luminosity would be consistent with the correlation found. Similarly, the IRS spectrum is too noisy for further analysis and it is unclear if the nucleus is detected here at all.

\subsection{\object{NGC 7626}} 
The {\it Chandra} observation of this object has only been published by \cite{humphrey_chandra_2006} but for the whole galaxy in the 0.1-10\,keV band. The [OIII] flux measured by \cite{ho_search_1997-1} predicts a much lower value ($\log L_\mathrm{X} = 39.56$) and will be used in this study. 
\cite{dudik_spitzer_2009} analyzed the forbidden emission lines seen by {\it Spitzer} and concluded that this object might not be an AGN based on the non-detection of [NeV] and broad H$\alpha$ emission. But the upper limit for the predicted X-ray luminosity ($\log L_\mathrm{X} < 41.03$) from the upper-limit  of the [OIV] non-detection is consistent with the [OIII] prediction. In addition, a point source was detected in radio \citep{nagar_radio_2005}. So the nuclear nature of NGC 7626 remains uncertain. The non-detection with VISIR favors a small X-ray luminosity but the upper-limit would be consistent with the MIR-X-ray correlation in any case.

\subsection{\object{NGC 7743}}
The properties of this object are very similar to NGC 5363: it might be either highly absorbed or not at all: based on {\it XMM-Newton} \cite{gonzalez-martin_x-ray_2009}  give an unobscured fit ($\log L_\mathrm{X} = 39.50$), while \cite{panessa_x-ray_2006}, \cite{gonzalez-martin_fitting_2009}, and \cite{akylas_xmm-newton_2009} interpret the X-ray data as Compton-thick. The [OIII] data  \citep[e.g.][]{gonzalez-martin_fitting_2009,panessa_x-ray_2006} is favoring Compton-thickness in agreement with [OIV] measurements stated in \cite{diamond-stanic_isotropic_2009}. Although there is no broad H$\alpha$ line detected \citep{terashima_hard_2000}, a compact radio core is present \citep{ho_radio_2001}. In summary, the true nature of this source is unclear but likely Compton-thick. Thus, the X-ray luminosity given by \cite{gonzalez-martin_x-ray_2009} will be used here. Note however, that a Compton-thin nature would be in better agreement with the MIR-X-ray correlation.

\end{document}